\documentclass[preprint]{elsarticle}

\usepackage{hyperref}
\hypersetup{
  colorlinks   = true, 
  urlcolor     = blue, 
  linkcolor    = blue, 
  citecolor   = blue 
}
\usepackage{lineno}

\journal{Journal of \LaTeX\ Templates}
\makeatletter
\def\ps@pprintTitle{%
    \let\@oddhead\@empty
    \let\@evenhead\@empty
    \def\@oddfoot{\footnotesize\itshape
         {Preprint} \hfill\today}%
    \let\@evenfoot\@oddfoot
    }
\makeatother

\setcitestyle{authoryear,open={(},close={)}} 

\usepackage{doi}

\usepackage{booktabs}
\usepackage{multirow}
\usepackage[table,xcdraw]{xcolor}
\usepackage{ulem}
\useunder{\uline}{\ul}{}
\usepackage{amssymb}
\usepackage{amsmath}
\usepackage[nameinlink]{cleveref}
\usepackage{caption}
\usepackage{subcaption}
\usepackage[export]{adjustbox}
\usepackage{bm}

\newcommand{\Lagr}{\mathcal{L}}
\newtheorem{definition}{Definition}
\newtheorem{problem}{Problem}

\newcommand{\beginsupplement}{%
        \setcounter{table}{0}
        \renewcommand{\thetable}{S\arabic{table}}%
        \setcounter{figure}{0}
        \renewcommand{\thefigure}{S\arabic{figure}}%
        \renewcommand{\thesubsection}{\Alph{subsection}}
     }

\begin{document}

\begin{frontmatter}

  \title{Context-aware multi-head self-attentional neural network model for next location prediction}


  \author[ikg]{Ye Hong\corref{correspondingauthor}}
  \cortext[correspondingauthor]{Corresponding author}
  \ead{hongy@ethz.ch}

  \author[frs,ikg]{Yatao Zhang}
  \ead{yatzhang@ethz.ch}

  \author[igp]{Konrad Schindler}
  \ead{schindler@ethz.ch}

  \author[ikg,frs]{Martin Raubal}
  \ead{mraubal@ethz.ch}

  \address[ikg]{Institute of Cartography and Geoinformation, ETH Zurich}
  \address[frs]{Future Resilient Systems, Singapore-ETH Centre, ETH Zurich}
  \address[igp]{Photogrammetry and Remote Sensing, ETH Zurich}

  \begin{abstract}
    Accurate activity location prediction is a crucial component of many mobility applications and is particularly required to develop personalized, sustainable transportation systems.
    Despite the widespread adoption of deep learning models, next location prediction models lack a comprehensive discussion and integration of mobility-related spatio-temporal contexts.
    Here, we utilize a multi-head self-attentional (MHSA) neural network that learns location transition patterns from historical location visits, their visit time and activity duration, as well as their surrounding land use functions, to infer an individual's next location.
    Specifically, we adopt point-of-interest data and latent Dirichlet allocation for representing locations' land use contexts at multiple spatial scales, generate embedding vectors of the spatio-temporal features, and learn to predict the next location with an MHSA network.
    Through experiments on two large-scale GNSS tracking datasets, we demonstrate that the proposed model outperforms other state-of-the-art prediction models, and reveal the contribution of various spatio-temporal contexts to the model's performance.
    Moreover, we find that the model trained on population data achieves higher prediction performance with fewer parameters than individual-level models due to learning from collective movement patterns. 
    We also reveal mobility conducted in the recent past and one week before has the largest influence on the current prediction, showing that learning from a subset of the historical mobility is sufficient to obtain an accurate location prediction result.
    We believe that the proposed model is vital for context-aware mobility prediction. The gained insights will help to understand location prediction models and promote their implementation for mobility applications.

  \end{abstract}

  \begin{keyword}
    Location prediction; Individual mobility; Context integration; Deep learning; Self-attention network
  \end{keyword}

\end{frontmatter}

\section{Introduction}

%
The movement of individuals is fundamental to our society~\citep{schlapfer2021universal} and an essential component to understanding many noticeable societal challenges, such as urbanization~\citep{barthelemy2019statistical}, urban activity~\citep{ahas_everyday_2015} and epidemic spread~\citep{kraemer_effect_2020}.
Central to the study of human mobility are the daily visited locations, where meaningful activities (such as working and shopping) are conducted~\citep{hong_conserved_2023}, and where individuals might interact to exchange knowledge~\citep{coscia_knowledge_2020}.
The ability to correctly predict the next activity location an individual will reach is required for many downstream tasks, such as personalized recommendation systems~\citep{sanchez_point_2022}, traffic optimization~\citep{rossi_modelling_2020}, mobile communication networks~\citep{zhang_mobility_2019}, and sustainable transportation systems~\citep{ma_individual_2022}.
Therefore, accurate location prediction is indispensable for mobility applications.

Despite its importance, location prediction is a challenging problem that is not yet fully tackled.
It is widely recognized that on the collective level, human movements exhibit a markedly regular pattern that can be described using statistical distributions~\citep{gonzalez_understanding_2008, song_modelling_2010}.
However, focusing on the individual level, mobility contains a lot of spontaneous decisions and exhibits complex patterns that influence the prediction performance~\citep{cuttone_understanding_2018}.
Mobility entropy~\citep{song_limits_2010} and related measures~\citep{chen_contrasting_2022, zhang_beyond_2022, yuan_analyzing_2016} proposed to measure the theoretical predictability of mobility suggest that specific sections of location visits are difficult to predict.
Another challenge to this problem falls on the highly context-dependent nature of mobility~\citep{buchin2012context}. Previous studies verify that the integration of context can increase movement predictability~\citep{zhang_beyond_2022, teixeira_deciphering_2019}.
However, the related context information, such as time, duration spent, and surrounding urban functions, is represented in diverse formats.
It is therefore challenging to summarize complex and high-dimensional relations between mobility and context as well as incorporate context into location prediction~\citep{xue2021mobtcast}.
%

The emergence of digital datasets containing human mobility traces and the development of deep learning (DL) models provide new opportunities to tackle mobility prediction problems.
Based on location-based services (LBS) and information and communication technologies (ICT)~\citep{huang_location_2018}, massive datasets recording individual positions with GNSS signals emerge, which include high-quality mobility diaries of individuals at a high temporal resolution~\citep{solomon_analyzing_2021}.
Compared to point-of-interests (POIs) check-in sequences generated by location-based social network (LBSN) applications, continuously recorded GNSS location traces contain complete information regarding an individual's activity schedules and mobility patterns~\citep{ma_individual_2022}. Thus, these datasets can uncover complex regularities in location visitation sequences.
Along with the widespread availability of digital datasets, DL models have been introduced for mobility prediction tasks~\citep{luca_survey_2021}.
These models are compelling in learning complex relations directly from data~\citep{manibardo_deep_2022, yan_learning_2022}.
In particular, models designed for sequence tasks, such as the long short-term memory (LSTM) network and the multi-head self-attention (MHSA)-based network, can learn periodicity and dependencies in an individual's location visits~\citep{feng_deepmove_2018}, which are properties challenging to be considered by traditional methods.
Moreover, the flexible architecture designs of DL models allow the integration of different data type formats; a process often referred to as data fusion~\citep{manibardo_deep_2022}. This flexibility ensures that context information related to mobility can be well combined and extracted for the location prediction task.

%
However, current mainstream approaches for next location prediction on GNSS tracking datasets only consider raw location sequences and their visit times, without properly discussing other contexts that might influence the performance.
In particular, time spent at locations is an indicator of conducted activities and might help identify an individual's daily routine~\citep{sun_joint_2021}.
Furthermore, abundant studies on travel behaviour have revealed the close relationship between human movements and the built environment, including the functional land use of locations~\citep{eldeeb_built_2021, losada_effect_2022}. Yet, these relevant contexts have not been sufficiently integrated into individual mobility prediction models.
Attempts to incorporate various context information utilizing MHSA networks have already started to gain attention for POI recommendation with LBSN check-in data~\citep{xue2021mobtcast}. Nevertheless, considering the discrepancies between check-in and GNSS tracking datasets, we anticipate differences in the relevant contexts and their effects on prediction.
Additionally, there is an ongoing discussion regarding modelling approaches for DL-based location prediction models, such as the length of the considered historical sequence~\citep{xu_understanding_2022} and the granularity of the prediction model~\citep{solomon_analyzing_2021}. These modelling choices heavily affect the prediction performance, and an in-depth understanding of their influence will benefit the design of more efficient models.

We aim to bridge the research gaps by proposing an MHSA-based model that integrates various context information for next location prediction.
Specifically, the model learns location transition patterns from raw location visit sequences, their temporal features (visitation time and activity duration), and their land use functions.
The latter is extracted with POI data and latent Dirichlet allocation (LDA), which generates functional land use semantics for each location at multiple spatial scales.
Moreover, through extensive experiments on two large-scale real-world GNSS tracking datasets, we analyse the influence of various modelling choices on prediction performance. We propose a set of optimum decisions for the task that help to better understand mobility prediction and guide the design and implementation of relevant models.
In short, our contributions are summarized as follows:
\begin{itemize}
  \item We design an MHSA-based model that integrates location features, temporal features, and functional land use contexts for next location prediction. The proposed context modelling approaches effectively capture movement-related spatio-temporal information, and the model achieves state-of-the-art performance on GNSS mobility datasets.
  \item We comprehensively discuss the model's granularity choice and prediction's historical dependencies. We find that the model trained on population data learns collective movement patterns and achieves higher prediction performance than individual-level models. We additionally show that movements of the recent past and weekly periodicity are vital for the prediction.
  \item We conduct extensive experiments to empirically show the effectiveness of our proposed framework and modelling choices. We open-sourced our framework for reproducing the results and providing a benchmark for further reference\footnote{The source code is available at \url{https://github.com/mie-lab/location-prediction}}.
\end{itemize}
The rest of this paper is organized as follows. We first systematically review related work in Section~\ref{sec:related}. In Section~\ref{sec:problem}, we formulate the next location prediction problem. Next, we present details of the prediction model in Section~\ref{sec:method}. Section~\ref{sec:exper} explains the tracking datasets and our experimental designs. The performance of our proposed model and the influence of various modelling choices are presented in Section~\ref{sec:result}. Finally, we summarize the main findings and conclude the paper in Section~\ref{sec:discussion}.

\section{Related work}\label{sec:related}

\subsection{Next location prediction methods}

Next location prediction can be broadly stated as predicting the immediate next location an individual will visit.
The exact definitions and methods to tackle the problem vary due to different objectives and the properties of employed datasets.
In particular, the problem can be identified in both recommendation systems and mobility behaviour studies. While the former focuses on predicting the following POI for LBSN applications~\citep{sanchez_point_2022}, the latter is interested in understanding where an individual will choose to conduct their next activity, typically dealing with data collected from GNSS signals~\citep{ma_individual_2022}.
Check-in datasets contain sequence visit information of many individuals, yet they only record a position when users manually check in to a pre-existing POI~\citep{sun_tcsa_2022}. In contrast, GNSS tracking studies continuously record individuals' whereabouts, but their participant number is often limited due to the high recruitment effort~\citep{solomon_analyzing_2021}.
Here, we focus on the methods proposed for GNSS-based datasets, as their mobility traces reflect individuals' full activity schedules, which are more helpful for transportation and mobility applications.

Studies on predicting individuals' next location have flourished over the last decade.
Before the DL explosion, Markov models were perhaps the most often employed methods for the task. \citet{AshbrookS02} and \citet{gambs_next_2012} identified significant locations from GNSS traces and built user-specific Markov models for next location prediction. These models regard locations as states and encode their pair-wise transition probability into a matrix.
To include personal and collective preferences, \citet{RendleFS10} proposed the factorized personalized Markov chains (FPMC) model for calculating the transition probability.
Later variants of models assuming the Markovian property of location visit further increased the prediction performance~\citep{huang_mining_2017, wang_predictability_2021}, yet they struggle to represent long-range spatio-temporal dependencies~\citep{li_hierarchical_2020}.
Therefore, DL models capable of learning complex sequential dependencies from a vast amount of data are introduced to tackle the task~\citep{luca_survey_2021}.
A typical example is the sequence modelling network LSTM, which was reported to outperform Markov models by a large margin for predicting the next location~\citep{krishna_lstm_2018, xu_understanding_2022}.
Still, LSTM suffers from the limited ability to relate information far apart when the length of the input sequences increases. Recent studies thus incorporate the attention mechanism with LSTM models to balance the weights assigned to recent and distant information, enabling the model to capture relevant temporal dependencies across the entire sequence~\citep{feng_deepmove_2018, li_hierarchical_2020}.
The success of attention has inspired the introduction of MHSA, which utilize several layers of self-attention to learn multiple sets of relations in the input sequence~\citep{Vaswani_2017}. This specific network architecture design brings more flexibility and capacity to capture the periodicity in individuals' location visits compared to LSTM models. 
MHSA-based models have already achieved state-of-the-art performances in POI recommendation tasks~\citep{xue2021mobtcast, sun_tcsa_2022}, but their potential for next location prediction on continuously recorded mobility traces is yet to be revealed.

Apart from predictive performance, the interpretability of location prediction models is a crucial aspect that influences their integration into real-world application systems~\citep{liu_dynamic_2021}. 
Model interpretability refers to the degree to which humans can understand the decision-making process of a model. 
For example, \citet{mo_individual_2022} proposed a hidden Markov model to predict the time and location of an individual's next trip, which can also be applied to analyze hidden activity patterns of individuals.
However, DL-based location prediction models, with their numerous interconnected parameters, are often criticized for their low interpretability, as it is challenging to reveal their inner workings~\citep{li_hierarchical_2020, ma_individual_2022}.
In this work, we enhance interpretability by analyzing the importance of input information and the learned attention weights from the proposed MHSA-based model, which reveal the information considered valuable for making predictions.

\subsection{Modelling approaches for next location prediction}
The specific details of how the next location prediction problem is implemented have a core impact on the prediction performance. Here, we refer to these as modelling approaches.
Two essential modelling choices for DL models in location prediction are the length of the considered historical sequences and the granularity of the model.
The sequence length determines how much information the DL model processes to infer a prediction~\citep{sun_tcsa_2022}. Its choice closely relates to the periodic nature of human movement~\citep{sun_understanding_2013}.
Since individual mobility exhibits both variability and stability over time~\citep{hong_conserved_2023}, increasing the input length allows the model to access more information, but might also include more irregular patterns that hinder the learning of stable location transitions.
Recent studies have started to discuss the impact of sequence length on the prediction performance with LSTM models~\citep{feng_deepmove_2018, xu_understanding_2022}. However, they often consider a fixed length for every input sequence, without distinguishing between users and the temporal context of the current prediction.
As abundant studies on travel behaviour have suggested regarding the day as the basic unit for mobility~\citep{schneider_unravelling_2013, dharmowijoyo_day_2016}, we propose to classify movement records based on their conducted day and quantify the impact of including sequences within different past days.

Whether to allow a prediction model to access movement records other than the current user is another modelling choice deserving careful consideration.
Although classical Markov models typically construct user-specific transition matrices~\citep{huang_mining_2017}, it has been shown that incorporating collective information leads to better prediction performance~\citep{cheng_where_2013}.
However, there is still an ongoing discussion of the granularity choice for DL models.
Collective-level models are more prevalent for check-in data because of their abundant user number that leads to highly overlapping POI sequences~\citep{xue2021mobtcast}.
The choice for models operating on GNSS mobility traces, where users rarely share a common location sequence in their daily mobility, is unclear.
While some studies proposed to train a single model for the whole population~\citep{krishna_lstm_2018, xu_understanding_2022}, several studies have shown that preserving a dedicated model for each user increases the prediction performance, with the cost of increasing parameters and training time~\citep{urner_assessing_2018, solomon_analyzing_2021}.
Here, we discuss the MHSA model's ability to infer an individual's next location choice from collective knowledge, providing evidence for the optimum choice of model granularity.

\subsection{Context dependence of human mobility}
Human mobility occurs in a situation-dependent contextual setting that affects the decision behaviours of individuals, which in turn impacts movement itself~\citep{sharif2017context}.
The broad scope of settings endorses the diversity of context data, covering various temporal and spatial information~\citep{sila2016analysis, gao2015spatio}.
From the temporal perspective, studies have discussed the potential influence of the day of the week, the time of the day and duration spent at different locations on individual movement~\citep{huang2015modeling, krishna_lstm_2018}.
For spatial context, \citet{buchin2012context} enumerated several examples in the urban environment, such as transportation networks, terrain, and land use.
Amongst the broad scope of spatial context, land use functions are closely related to human mobility~\citep{lee2015relating}. POI data demonstrate great potential to represent land use due to their high availability and rich semantic information~\citep{hong_hierarchical_2019, tu2020scale}.
Initially, counting the distribution of POI, e.g., using the term frequency-inverse document frequency (TF-IDF) method~\citep{yuan2014discovering}, was proposed to identify urban functions. However, this method fails to exploit the abundant multi-categorical information of POIs~\citep{liu2017classifying}.
To solve this problem, \citet{zhang2019functional} utilized probabilistic topic modelling to generate semantic representations of land parcels, improving the accuracy of functional land use recognition.
As a representative example of context information, modelling POI has received much attention and has already found applications in various transport tasks, including safety analysis~\citep{jia_traffic_2018} and traffic prediction~\citep{krause_short_2019}. However, specific approaches for incorporating POI data into models for location prediction are currently lacking, leaving the potential benefits of considering the POI context unexplored.

Most existing studies only use raw location visit sequences of individuals to predict their next locations, regardless of the current mobility context~\citep{laha2018real}. This is due to the complexity and diversity of context data~\citep{tedjopurnomo2020survey}.
With the development of machine learning, studies have started to explore methods for context data fusion and employ them for boosting human mobility research~\citep{zheng2018survey, lau2019survey}.
However, context data from multiple sources possess complex formats, hindering their fusion into a unified representation~\citep{liao2018multi}.
Context-aware DL frameworks provide a solution to combine multi-source context information~\citep{sun_tcsa_2022}, yet they most often focus on the temporal aspect and overlook complex spatio-temporal interactions.
To further explore the influences of context information, we propose a novel MHSA-based model to incorporate the spatial and temporal context into the next location prediction problem.

\section{Problem definition}\label{sec:problem}
We introduce a set of terms and notions used in the remainder of the article and formulate the next location prediction problem.

Mobility data are typically collected through electronic devices and stored as spatio-temporal trajectories. Each track point in a user's trajectory contains a pair of spatial coordinates and a timestamp.

\begin{sloppypar}
  \begin{definition}[GNSS Trajectory]
    Let $u^i$ be a user from the user set $\mathcal{U} = \left \{ u^{1}, ..., u^{\left |\mathcal{U}  \right |} \right \}$, a trajectory $T^i = \left (q_k  \right )_{k=1}^{n_{u^i}}$ is a time-ordered sequence composed of $n_{u^i}$ track points visited by $u^i$.  A track point can be represented as a tuple of $q = \langle p, t\rangle $, where $p = \langle x, y\rangle $ represents spatial coordinates in a reference system, e.g., latitude and longitude, and $t$ is the time of recording. 
  \end{definition}
\end{sloppypar}

Stay points are detected from raw GNSS trajectories to identify areas where users are stationary for a minimum amount of time~\citep{li_mining_2008}.
Then, locations are formed through spatially aggregating stay points for characterizing place semantics, such as the functional land use or opening hours of a shop~\citep{hariharan2004project, Martin_trackintel_2023}. We present an example of this movement data generation process in Appendix~\ref{app:movement}.

\begin{definition}[Stay point]
  A Stay point $S = \left (q_k  \right )_{k=start}^{end} $ is a subsequence of Trajectory $T^i$ where user $u^i$ was stationary from the start track point $q_{start}$ to the end track point $q_{end}$. Each stay point $S$ can be represented as a tuple of $\langle t, d, g(s) \rangle$, where $t$ and $d$ represent the start timestamp and the stayed duration, respectively, and $g(s)$ denotes the geometry, often represented as the centre of its containing track points. We use $S_k$ to denote the $k$-th stay point in a user's GNSS trajectory.
\end{definition}

\begin{definition}[Location]
  A location $L$ consists of a set of spatially-proximate stay points. It can be represented as a tuple $L = \langle l, c, g(l) \rangle$, where $l$ is the location identifier, $c$ encodes its context semantics such as the surrounding land use, and $g(l)$ denotes the geometry of the location, calculated as the convex hull of all contained stay points. Therefore, each location is defined as an area. We define $\mathcal{O}^i$ as the set containing known locations for user $u^i$, and $\mathcal{O} = \left \{ \mathcal{O}^1, ..., \mathcal{O}^{\left |\mathcal{U}  \right |} \right \}$ as the set containing all locations.
\end{definition}

Through the generation of locations, each stay point is enriched with its context semantics, i.e., $S = \langle t, d, g(s), l, c, g(l) \rangle$, and a user's mobility can be represented as a time-ordered sequence of $N$ visited stay points $\left (S_k  \right )_{k=1}^{N}$. Next, we define the next location prediction problem.

\begin{problem}[Next location prediction]
Consider a stay point sequence with context information $\left (S_k  \right )_{k=m}^{n}$ visited by user $u^i$ in a time window from time step $m$ to $n$, the goal is to predict the location the same user will visit in the next time step, i.e., the location identifier $l_{n+1} \in \mathcal{O}$.
\end{problem}

The time window length determines how much historical information is considered in the predictive model. Here, we construct the historical sequence based on mobility conducted in the past $D \in \left \{ 0, 1, ..., 14 \right \}$ days\footnote{$D=0$ indicates we only consider stay point sequences visited in the current day.}. Hence, the historical window length depends on the user $u^i$ and the current time step $n$. Next location prediction is defined as a sequence prediction problem with variable sequence lengths.

\section{Methodology}\label{sec:method}
\begin{figure}[htb!]
  \centering
  \includegraphics[width=0.9\linewidth]{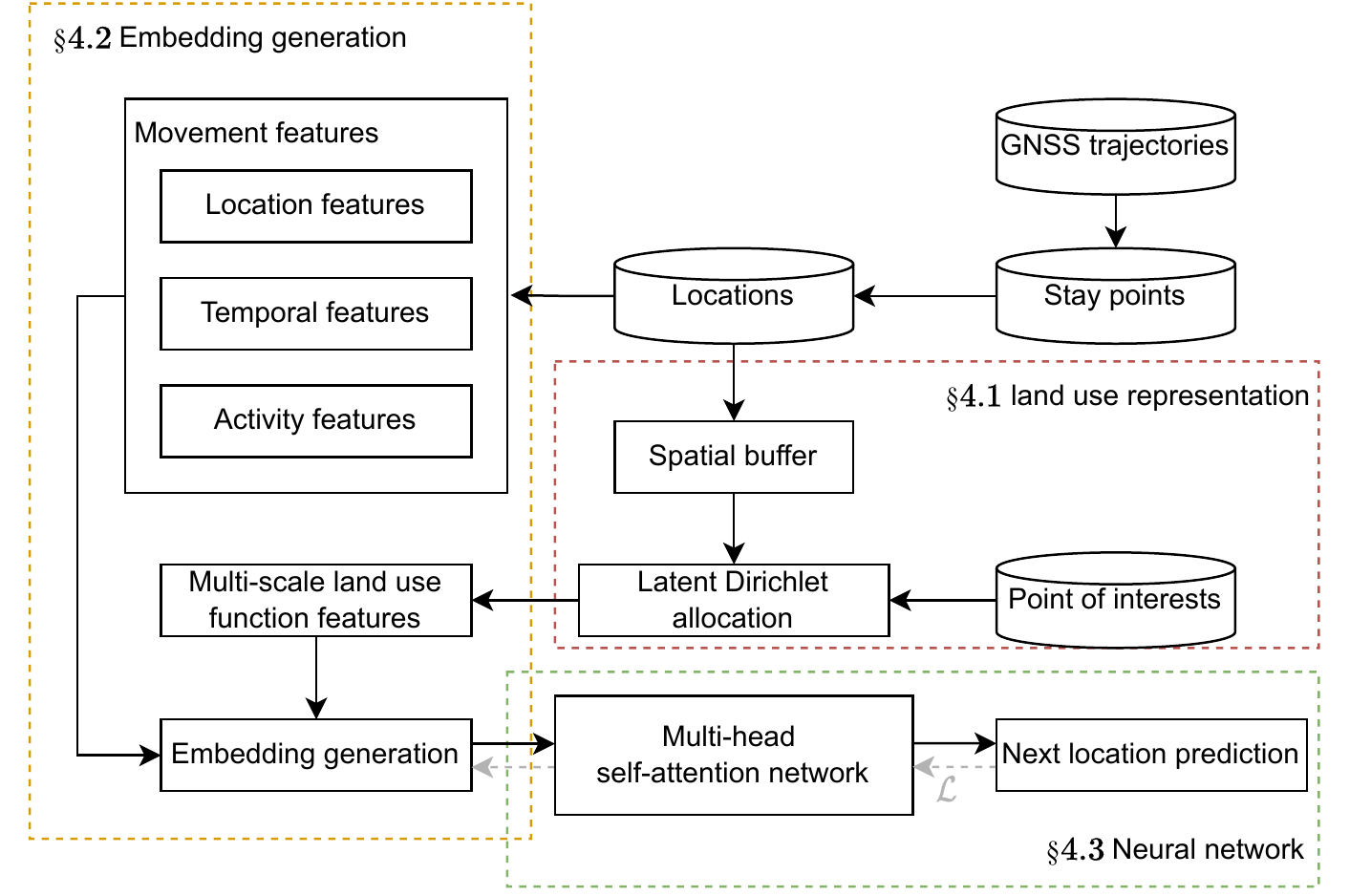}
  \caption{Overview of the proposed framework for context-aware next location prediction. We represent the land use context of each location with point-of-interests (POIs) and latent Dirichlet allocation (LDA) ($\S$\ref{sec:land}), generate the embedding vectors of spatio-temporal features ($\S$\ref{sec:emb}), and learn to predict the next location with a multi-head self-attention (MHSA) network ($\S$\ref{sec:network}).}
  \label{fig:overview}
\end{figure}

We propose a neural network that utilizes context information to tackle next location prediction. The overall pipeline is depicted in Figure \ref{fig:overview}.
First, we represent the land-use context around each location using POI data ($\S$\ref{sec:land}). The mixtures of land use functions are measured using LDA at different spatial scales.
Then, we utilize various embedding layers to represent the heterogeneous movement and context data ($\S$\ref{sec:emb}).
Finally, we adapt the MHSA network to learn the dependencies from the historical sequence and infer the next visiting location ($\S$\ref{sec:network}). In the following, we provide a detailed description of each module.

\subsection{Representing land use context}\label{sec:land}
\begin{figure}[htb!]
  \centering
  \includegraphics[width=0.9\linewidth]{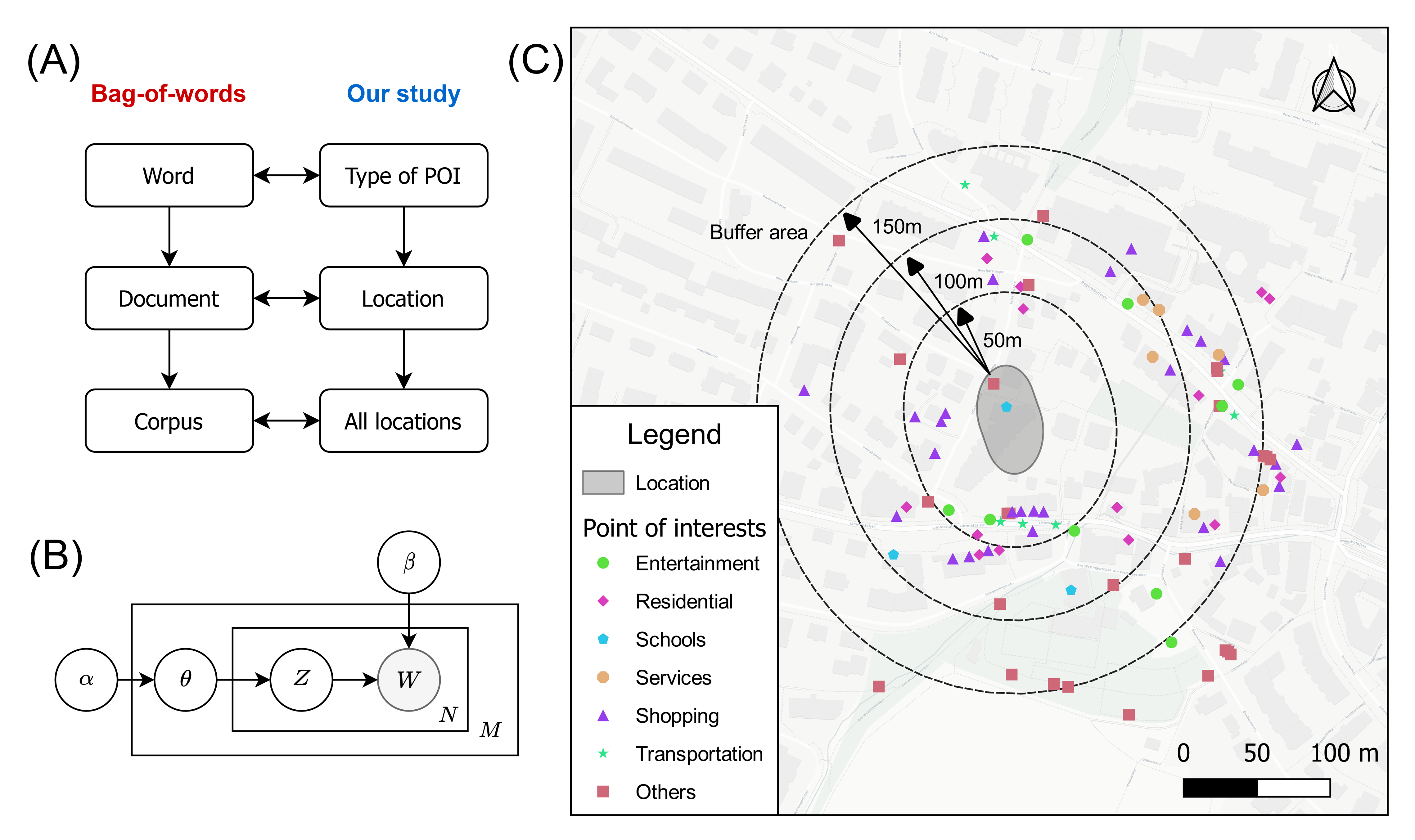}

  \caption{Land use context representation for locations. (A) Concepts in the bag-of-words (BOW) model and their correspondence in this study. (B) Graphical model of LDA that is utilized to learn the latent topic distribution of each location using a bag-of-POIs representation. (C) Multi-scale bag-of-POIs are constructed by creating buffer areas with various distances around each location. Map data ©OpenStreetMap contributors, ©CARTO. }
  \label{fig:lda}
\end{figure}

We start by encoding the semantics of locations through their surrounding land use information, which serves as a proxy for describing individuals' activities and is valuable for uncovering preferences in location choices~\citep{calabrese_human_2010}. To extract the land use function semantics, we employ probabilistic topic modelling methods, specifically the bag-of-words (BOW) and LDA models. Initially proposed for text modelling in natural language processing, the BOW method represents a \textit{document} as a multiset of its \textit{words}, disregarding word order but considering word frequency. BOW also regards the set of all considered documents as a \textit{corpus} (Figure~\ref{fig:lda}A). Next, we apply the LDA model to the corpus to generate a semantic representation for each document based on its multisets of words. This representation is a mixture of latent topics, each characterized by word distributions~\citep{blei_latent_2003}. The graphical model of LDA, shown in Figure~\ref{fig:lda}B, is a three-level generative model with $M$ documents and $N$ words. Compared to one-hot embedding and TF-IDF methods, LDA generates dense vector representations to characterize document semantics and has now become a standard approach to represent POIs in the geographical space~\citep{zhang2019functional}. 

More specifically, we regard POI categories as words, locations as documents, and the set of all locations as the corpus (Figure~\ref{fig:lda}A). We first construct an area with a buffer size $dist$ around each location to consider its surrounding POIs in the urban space (Figure~\ref{fig:lda}C). Given $P$ categories of POIs, each location $L_j$ can be represented with its bag-of-POIs using a vector representation $b_j$:
\begin{linenomath}
  \begin{equation}
    b_j = [w_1, ..., w_i, ..., w_P]^T
  \end{equation}
\end{linenomath}
where $w_i$ represents the number of times the $i$-th POI category appears within the range $dist$ to the location. We obtain the location-POI frequency matrix $\boldsymbol{B}$ through stacking all location's bag-of-POIs, i.e.,  $\boldsymbol{B}=[b_1, ..., b_j, ..., b_{\left |\mathcal{O}  \right |} ]$. $\boldsymbol{B}$ can be further processed with LDA to infer the latent topic distribution $c_j(dist) \in \mathbb{R}^{tp}$ for each location $L_j$, where $tp$ represents the number of topics that is pre-specified for LDA.

We propose to represent the land use context at multiple spatial scales. The dependence between land use and mobility varies across space, meaning a fixed spatial range is insufficient to encode land use for locations in different geographic areas~\citep{zhang2022street}. Therefore, we alter the buffer range $dist$ to control the inclusion of POIs, allowing us to capture spatial characteristics at different scales by limiting the influence of POIs on locations within a specific range. See Figure~\ref{fig:lda}C for an example of selecting 3 different buffer distances around a location.
With the same topic identification process, we now acquire an array of semantic representation vectors for each location.
Instead of manually selecting the appropriate scales, we propose to retain the multi-scale information and extract the most relevant ones for location prediction using a neural network. Hence, the context information matrix $\boldsymbol{C_j}$ is a stack of all vectors obtained for location $L_j$:
\begin{linenomath}
  \begin{equation}
    \boldsymbol{C_j} = [c_j(dist_1), ..., c_j(dist_i), ..., c_j(dist_K)]
  \end{equation}
\end{linenomath}
where $K$ is the total number of considered buffer distances and $c_j(dist_i)$ denotes the semantic vector obtained through choosing $dist_i$ to create the bag-of-POIs. In this study, we consider $K=11$ buffer distances that increase by 50 meters per step, i.e., $dist \in [0, 50, 100, ..., 500]$ meter. As a result, each location's multi-scale land use context is encoded in $\boldsymbol{C}$, which can be further processed with a neural network.

\subsection{Generating spatio-temporal embedding}\label{sec:emb}

An accurate location prediction model requires appropriate selection and modelling of historical sequence information.
Besides the often included raw location identifier and the corresponding time of visit~\citep{li_hierarchical_2020}, we consider the activity duration and land use functions for describing each visited stay point, which ensures a comprehensive representation of its context from a spatio-temporal perspective.
Moreover, user-related information helps to uncover sequences travelled by different users and assists the network in learning user-specific movement patterns.
We use embedding layers to represent features from the categorical type to a real-valued vector. As opposed to the more classical one-hot representation, embedding vectors are more compact and can effectively capture the latent correlation between different feature types~\citep{xu_understanding_2022}. These layers are parameter matrices that provide mappings between the original variable and the embedding vector, jointly optimized with the entire network.
The process of generating the spatio-temporal embedding is shown in Figure~\ref{fig:network}. The details of the embedding layers and the context network can be found in Appendix~\ref{app:network}.

\begin{figure}[htbp!]
  \centering
  \includegraphics[width=0.8\linewidth]{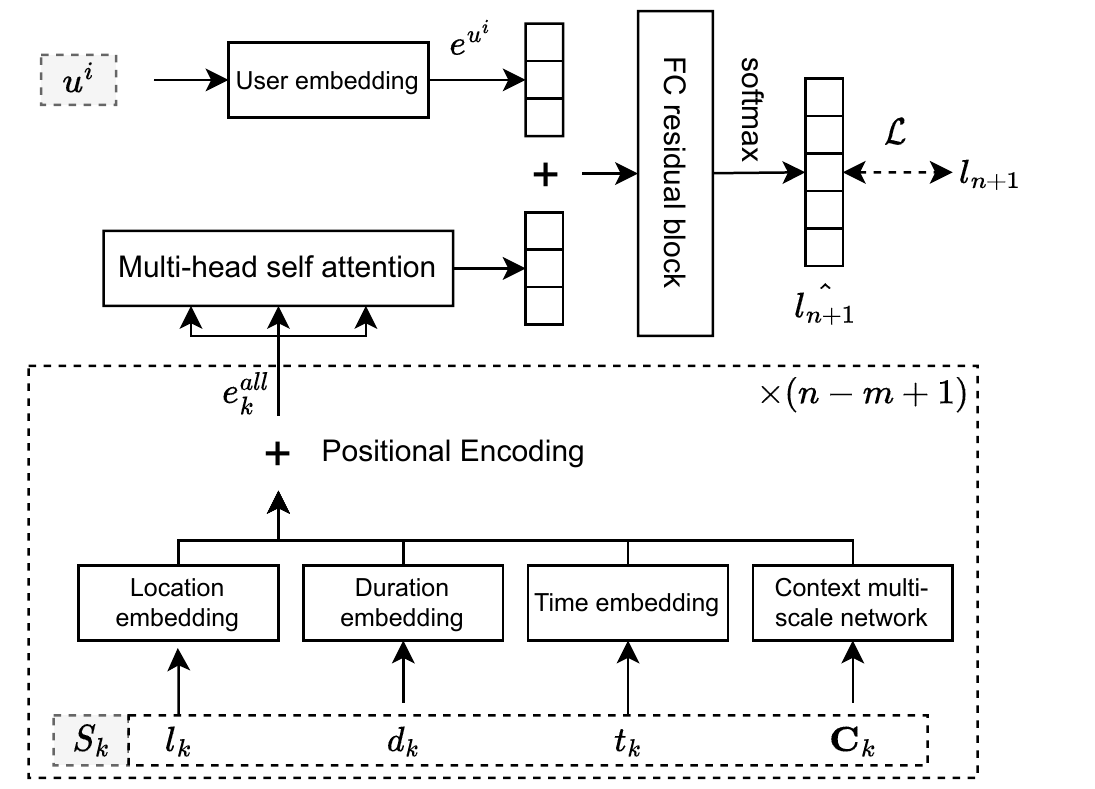}
  \caption{Embedding layers and the MHSA-based network for next location prediction. Spatio-temporal features at each time step are transformed into embedding vectors. These vectors are combined and fed into the MHSA network, whose output vector is combined with the user embedding to generate a location prediction.}
  \label{fig:network}
\end{figure}

Operationally, given a stay point $S_k$ in the historical sequence, its location identifier $l_k$, time of arrival $t_k$, and spent duration $d_k$ are fed into their respective embedding layers to generate vector representations:
\begin{linenomath}
  \begin{equation}
    e^{l}_{k} = h^l(l_{k}; \boldsymbol{W}^l) \qquad e^{t}_{k} = h^t(t_{k}; \boldsymbol{W}^{t}) \qquad e^{d}_{k} = h^d(d_{k}; \boldsymbol{W}^{d})
  \end{equation}
\end{linenomath}
where $e^{l}_{k}$, $e^{t}_{k}$ and $e^{d}_{k}$ are the respective embedding vectors, and $l_{k},\:t_{k},\:d_{k}$ are the respective categorical features. $h(\cdot; \cdot)$ denotes the embedding operation and $\boldsymbol{W}$ terms are the parameter matrices optimized during training.
We separately embed the minute, the hour, and the day of the week from the time of arrival $t_k$ to capture different levels of periodicity in historical visits.

To learn from the land-use context matrix $\boldsymbol{C}_k$, we apply a small context network $f^{\boldsymbol{C}}(\cdot; \cdot)$ that consists of two blocks of linear layers with residual connection and layer normalization, separated by a ReLU activation and a dropout layer.
Finally, the overall embedding vector $e^{all}_{k}$ for stay point $S_k$ is obtained by adding its features together with a positional encoding $PE$ that encodes the sequence information $k$:
\begin{linenomath}
  \begin{equation}
    e^{all}_{k} = e^{l}_{k} + e^{t}_{k} + e^{d}_{k} + f^{\boldsymbol{C}}(\boldsymbol{C}_k; \boldsymbol{W}^{\boldsymbol{C}})  + PE
  \end{equation}
\end{linenomath}
We use the original positional encoding proposed by~\citet{Vaswani_2017} that utilizes sine and cosine functions. The involvement of positional encoding is essential for training a self-attention network as it does not implicitly assume the sequential order for its input~\citep{Vaswani_2017}. Additionally, we represent the user \(u^i\) from which the stay point sequence is recorded into a vector \(e^{u^i}\) with a user embedding layer, i.e., $e^{u^i} = h^u({u^i}; \boldsymbol{W}^{u})$. Involving user information ensures that a model trained on population data can still distinguish traces travelled by different users, which we will discuss in detail in Section~\ref{sec:result}. As a result, we obtain the overall embedding vector $e^{all}_{k}$ that encodes spatio-temporal features at each time step and the user embedding vector $e^{u^i}$ for the sequence.

\subsection{Multi-head self-attention network}\label{sec:network}
Once we acquire dense spatio-temporal feature vectors at each time step, we must mine their sequential transition patterns. These historical patterns are captured utilizing an MHSA-based network, a mechanism originally proposed within the transformer network to tackle language translation tasks~\citep{Vaswani_2017}. We adopt an architecture similar to the Generative Pre-trained Transformer (GPT) that only includes the decoder part of the transformer model~\citep{radford2018improving}. The decoder consists of a stack of $L$ identical blocks, each with two components. The first is the masked multi-head self-attention, and the second is a feedforward network with two linear layers, separated by a ReLU activation function. Residual connections, layer normalization, and dropout layers are added to each component to facilitate learning. Details of the architecture can be found in Appendix~\ref{app:network}.

The output of the MHSA model $out_{n}$ is added with the user embedding $e^{u^{i}}$, and together fed into a fully-connected (FC) residual block. Finally, the predicted probability of each location is obtained through a softmax transformation:
\begin{linenomath}
  \begin{equation}
    \label{equation:fc}
    P(\hat{l}_{n+1}) = \text{Softmax(} f^{FC}(out_{n} + e^{u^{i}}; \boldsymbol{W}^{FC}) )
  \end{equation}
\end{linenomath}
where $f^{FC}(\cdot; \cdot)$ represents the FC residual block operation. This block consists of linear layers with residual connections, aiming at learning dependencies between the sequence information and the user for extracting personal mobility preferences. $P(\hat{l}_{n+1}) \in \mathbb{R}^{\left | \mathcal{O}  \right |}$ contains the probability of all locations to be visited at the next time step.

During training, with access to the ground truth next location $l_{n+1}$, the task can be regarded as a multi-class classification problem. Parameters of the model can therefore be optimized using the multi-class cross-entropy loss $\Lagr$:
\begin{linenomath}
  \begin{equation}
    \label{equation:loss}
    \Lagr = -\sum_{k=1}^{\left |\mathcal{O}  \right |}P(l_{n+1})^{(k)}\log(P(\hat{l}_{n+1})^{(k)})
  \end{equation}
\end{linenomath}
where $P(\hat{l}_{n+1})^{(k)}$ represents the predicted probability of visiting the $k$-th location and $P(l_{n+1})^{(k)}$ is the one-hot represented ground truth, i.e., $P(l_{n+1})^{(k)}=1$ if the ground truth next location is the $k$-th location, and $P(l_{n+1})^{(k)}=0$ otherwise.

\section{Experiment}\label{sec:exper}

\subsection{Data and preprocessing}

We demonstrate the effectiveness of the proposed method through extensive experiments on two longitudinal GNSS tracking datasets.

\textit{GC dataset}. The \textit{Green Class} (GC) dataset is an outcome of the SBB Green Class E-Car pilot study conducted by the Swiss Federal Railways (SBB) from November 2016 to December 2017. The study aims to evaluate the effect of a Mobility-as-a-Service (MaaS) offer on individuals' mobility behaviour~\citep{martin_begleitstudie_2019}. In the study, 139 participants were provided with a MaaS bundle and were asked to install a GPS-tracking application (app) on their smartphones that records their daily movement, with a median time of 13.9 seconds between two consecutive GPS recordings. The app pre-processed the raw GPS traces to infer \textit{stay points} where users are stationary and \textit{stages} of continuous movements that use a single travel mode. Analysis of the socio-demographic information suggests that the dataset is biased towards middle- and upper-class people with high mobility demands, most likely caused by participation preconditions~\citep{martin_begleitstudie_2019}.

\textit{Geolife dataset}. The \textit{Geolife} dataset released by Microsoft Research Asia was collected in the Geolife project involving 178 users in China from April 2007 to October 2011~\citep{zheng2010geolife}. The dataset contains a broad range of users' outdoor movements, including daily routines, such as going home and travelling to work, as well as entertainment and sports activities, such as shopping and sightseeing. These trajectories were recorded by different GPS loggers and phones with various sampling rates. 91\% of the trajectories are logged in a dense representation, i.e., every $1-5$ seconds or every $5-10$ meters per point. Each GPS trajectory in this dataset is represented by a sequence of time-stamped geocoordinate pairs. We follow the framework proposed in~\citet{zheng2010geolife} and classify the raw GPS trajectories into stationary \textit{stay points} and movement \textit{stages} using the \textit{Trackintel} library~\citep{Martin_trackintel_2023}.

The raw movement traces data from GNSS tracking studies are pre-processed for the next location prediction. We pre-filter the datasets to only consider users observed for more than 300 days in GC and more than 50 days in Geolife to ensure a long observational time. Moreover, we use temporal tracking coverage, which quantifies the proportion of time the users' whereabouts are recorded, to evaluate the tracking quality in the temporal dimension. After this process, 93 users in GC and 45 users in Geolife remain.
Furthermore, locations are generated from the individual visited stay point sequence. We regard a stay point as an \textit{activity} if its duration is longer than 25 minutes. Then, the activity stay points are spatially aggregated into \textit{locations} to account for visits to the same place at different times. We utilized the function provided in \textit{Trackintel} with parameters \(\epsilon = 20\) and \(num\_samples = 2\) to generate \textit{dataset} locations~\citep{hong_clustering_2021}. Table~\ref{tab:number} shows basic statistics for both considered datasets. Further mobility indicators revealing dataset properties are shown in Appendix~\ref{app:data}.

\begin{table}[htbp!]
  \centering

  \caption{Basic statistics of the  mobility datasets. The mean and standard deviation across users are reported.}
  \label{tab:number}
  \begin{tabular}{@{}ccc@{}}
    \toprule
                           & GC             & Geolife       \\ \midrule
    User number            & 93             & 45            \\
    Tracking period (days) & $418 \pm 19$   & $345 \pm 413$ \\
    \#Stay Points per user & $1398 \pm 399$ & $369 \pm 456$ \\
    \#Stay Points per user day & $3.9 \pm 2.3$ & $2.4 \pm 1.5$ \\
    \#Locations per user   & $179 \pm 49$   & $77 \pm 108$  \\
    Location size (m\textsuperscript{2}) & $1365 \pm 2389$          & $3606 \pm 12275$              \\
    Tracking coverage (\%) & $93 \pm 5$     & $44 \pm 24$   \\ \bottomrule
  \end{tabular}
\end{table}

The POI dataset utilized in this study derives from OpenStreetMap (OSM) (\url{http://www.openstreetmap.org}), an open-source project that provides users with free and easily accessible digital map resources~\citep{hong_hierarchical_2019}. We acquire historical POI data in Switzerland from early 2017 to match the time frame of the GC tracking study. We do not include POI data for the Geolife dataset, as the coverage and quality of the OSM data in China were relatively low around 2010~\citep{zheng_assessing_2014}. Therefore, no land use context information is considered for the Geolife dataset. The Switzerland POI dataset is supplemented with the 2017 building data from OSM. For each building data entry, we preserve its attribute values but regard the centre point of its polygon geometry as its new geometry, i.e., a new POI. The final POI dataset includes 987,866 POIs, grouped into 22 first-level and 404 second-level categories. We regard the second level category as the functional type description for each POI and adopt it in the process of generating land use representations for locations.

\subsection{Model training}

We split the GNSS tracking datasets into non-overlapping train, validation, and test sets with the ratio of 6:2:2 based on time, such that stay point sequences occurred in the first 60\% of tracking days for each user are regarded as train and the last 20\% of days as test. The parameters of the prediction network are fitted on the training set. The validation set is not used in the optimization but serves to monitor the network loss. We conduct a grid search on the validation set over the hyper-parameters. The detailed ranges and the final selected hyper-parameter set are given in Appendix~\ref{app:network}. We finally evaluate and report the model performances using the held-out test set.

During training, we minimize Eq.~\ref{equation:loss} with the Adam optimiser over batches of training data samples, with an initial learning rate of $1e^{-3}$ and an L2 penalty of $1e^{-6}$. An early stopping strategy is adopted to pause the learning if the validation loss stops decreasing for 3 epochs. Then, the learning rate is multiplied by 0.1, and training is continued from the model with the lowest validation loss. This early stopping process is repeated 3 times. Additionally, we implement learning rate warm-up for 2 epochs and a linear decay of 0.98 per epoch afterwards~\citep{Vaswani_2017}.

\subsection{Baseline prediction models}
We compare the performance of our proposed model with classical prediction methods and DL-based models. For each DL model, we keep the input features the same as our proposed network such that the differences in performance are only due to the variations in the model's architecture. 
The parameter number and computation time of each implemented model can be found in Appendix~\ref{app:network}.

\begin{itemize}
  \item Markov. Classical location prediction models assume the Markovian property for individual location visits~\citep{AshbrookS02}. We implement the first-order Mobility Markov Chain (1-MMC)~\citep{gambs_next_2012}, as increasing the order does not improve the prediction performance.
  \item FPMC. Originally proposed by \citet{RendleFS10} for next basket item recommendation, FPMC has been successfully applied to POI recommendations for check-in data~\citep{cheng_where_2013}. FPMC builds personalized Markov transition matrices and factorizes them jointly with a pairwise interaction model.
  \item LSTM. As a classical DL architecture for sequence modelling, LSTM is widely adopted and regarded as one of the state-of-the-art models for next location prediction~\citep{krishna_lstm_2018, solomon_analyzing_2021, xu_understanding_2022}. LSTM models retain a common hidden state that sequentially operates over the inputs, effectively capturing sequential relations.
  \item LSTM with self-attention (LSTM attn). Inspired by~\citet{li_hierarchical_2020} and the effectiveness of the attention module in Transformer, we implemented an LSTM with a (masked) self-attention between the hidden states. The attention results are combined with the actual output of the LSTM through a small feedforward network.
  \item DeepMove. The framework proposed by~\citet{feng_deepmove_2018} predicts the next location of an individual from their sparsely sampled trajectories. DeepMove consists of two separate recurrent networks, one for capturing the periodicity from historical visits and the other for mining transition patterns from the current trajectory. Their results are combined through an attention layer to infer the final prediction.
  \item MobTcast. Based on the transformer encoder network, \citet{xue2021mobtcast} proposed MobTCast to predict the next POI. MobTcast leverages temporal, semantic, social, and geographical contexts in the history POI sequence. It also introduces an auxiliary task to encourage predicting geographically proximate POIs. We do not include the social context since the overlap between location sequences of our users is low.
\end{itemize}

\subsection{Evaluation metrics}
We use the following metrics to quantify the performance of the implemented models:
\begin{itemize}
  \item Accuracy. It measures the correctness of the predicted location compared to the ground truth of the next visited location. Practically, we rank the location probability vector $P(\hat{l}_{n+1})$, obtained from Eq.~\ref{equation:fc}, in descending order and check whether the ground truth location appears within the top-k predictions. Acc$@$k measures the proportion of times this is true in the test dataset. In location prediction literature, this metric is also referred to as Recall$@$k or Hit Ratio$@$k. We report Acc$@$1, Acc$@$5, and Acc$@$10 to allow comparisons with other work.
  \item F1 score (F1). Individual visits to locations are highly unbalanced, with specific locations occurring more often in the daily schedule than others. We use the F1 score weighted by the visitation number to emphasize the model's performance in the more important locations.
  \item Mean reciprocal rank (MRR). It calculates the average rank reciprocal at which the first relevant entry was retrieved in the prediction vector:
        \begin{linenomath}
          \begin{equation}
            \label{equation:MRR}
            \text{MRR}={\frac  {1}{N}}\sum _{{i=1}}^{{N}}{\frac  {1}{{\text{rank}}_{i}}}
          \end{equation}
        \end{linenomath}
        where $N$ denotes the number of test samples and $\text{rank}_{i}$ is the rank of the ground truth location in $P(\hat{l}_{n+1})$ for the $i$-th test sample.
  \item Normalized discounted cumulative gain (NDCG). It measures the quality of the prediction vector by the ratio between the discounted cumulative gain (DCG) and the ideal discounted cumulative gain (IDCG):
    \begin{linenomath}
      \begin{equation}
        \label{equation:NDCG}
        \text{NDCG}=\frac{1}{N}\sum_{i=1}^{N}{\frac{\text{DCG}_i}{\text{IDCG}_i}}, \quad\text{where}\;\;\text{DCG}_i=\sum_{j=1}^{\left |\mathcal{O}  \right |}\frac{r_j}{\log_2(j+1)}
      \end{equation}
    \end{linenomath}
    where $r_j$ denotes the relevance value at position $j$. In the context of location prediction, $r_j$ is binary, that is, $r_j \in \{0, 1\}$, and $r_j=1$ if and only if the $j$-th item in the ranked $P(\hat{l}_{n+1})$ corresponds to the ground truth next location. NDCG$@$k measures the relevance of the results up to rank position k. In our evaluation, we report NDCG$@$10.
\end{itemize}

\section{Results}\label{sec:result}

\subsection{Performance results}

We first present the prediction performance for all considered methods in Table~\ref{tab:overall}. For each learning-based model, we train the model five times with different random parameter initialization and report the respective performance indicators' mean and standard deviation. We use the Mann-Whitney U test to check whether the performance gaps between different models are significant. DL models are trained on the whole population data, with user identifiers to distinguish sequences recorded from different users. We input historical sequences conducted in the past $D=7$ days for all DL models to guarantee their comparability. 

%

\addtolength{\tabcolsep}{-2pt}
\begin{table*}[htbp!]
\centering
\caption{Performance evaluation results for next location prediction. The mean and standard deviation across five different runs are reported. Numbers marked in \textbf{bold} and \underline{underline} represent the best and second-best performing method, respectively. Mann-Whitney U test is used to show whether performances are statistically different. }
\label{tab:overall}

    \makebox[\textwidth]{
    \begin{tabular}{@{}cccccccc@{}}
    \toprule
    Dataset & Method    &  Acc@1 & Acc@5 & Acc@10 & F1  & MRR & NDCG@10      \\ \midrule
    \multirow{6}{*}{GC} & 1-MMC     & 32.8 & 51.8 & 54.7 & 26.4 & 41.3 & 44.4 \\
                        & FPMC      & $34.2 \pm 0.9$ & $61.6 \pm 0.1$ & $66.6 \pm 0.3$ & $18.6 \pm 0.7$ & $46.9 \pm 0.4$ & $51.0 \pm 0.2$\\
                        & LSTM      & $43.6 \pm 0.2$ & \underline{$64.2 \pm 0.1$} & \underline{$69.0 \pm 0.1$} & $32.2 \pm 0.3$ & $53.1 \pm 0.1$  & $56.7 \pm 0.1$\\
                        & LSTM attn & $44.5 \pm 0.2$ & \underline{$64.3 \pm 0.1$} & \underline{$68.9 \pm 0.2$} & \underline{$33.2 \pm 0.4$} & \underline{$53.5 \pm 0.1$}  & \underline{$57.0 \pm 0.1$}\\
                        & Deepmove  & $43.2 \pm 0.1$ & $63.8 \pm 0.1$ & $68.4 \pm 0.2$ & $32.1 \pm 0.1$ & $52.6 \pm 0.1$ & $56.2 \pm 0.1$\\
                        & MobTcast  & \underline{$44.8 \pm 0.1$} & $63.8 \pm 0.2$ & $68.7 \pm 0.2$ & \underline{$33.6 \pm 0.2$} & \underline{$53.5 \pm 0.1$} & $56.8 \pm 0.1$\\
                        & Ours (MHSA)     & $\bm{45.6 \pm 0.1}$ & $\bm{64.7 \pm 0.1}$ & $\bm{69.5 \pm 0.1}$ & $\bm{34.8 \pm 0.2}$ & $\bm{54.3 \pm 0.1}$ & $\bm{57.6 \pm 0.1}$\\ \midrule

    \multirow{6}{*}{Geolife}        
                        & 1-MMC     & 24.1 & 38.1 & 39.5 & $\bm{22.7}$ & 30.5 & 32.7 \\
                        & FPMC      & $24.0 \pm 0.6$ & $53.7 \pm 2.0$ & $57.8 \pm 0.4$ & $13.5 \pm 0.6$ & $35.5 \pm 0.7$ & $40.8 \pm 0.6$ \\
                        & LSTM      & $28.4 \pm 0.8$ & $\bm{55.8 \pm 1.3}$ & \underline{$59.1 \pm 0.7$} & $19.3 \pm 0.8$ & \underline{$40.2 \pm 1.1$} & \underline{$44.7 \pm 0.6$}\\
                        & LSTM attn & \underline{$29.8 \pm 0.7$} & $\bm{54.6 \pm 1.5}$ & \underline{$58.2 \pm 1.7$} & \underline{$21.3 \pm 0.8$} & \underline{$40.7 \pm 0.4$} & \underline{$45.0 \pm 0.7$}\\
                        & Deepmove  & $26.1 \pm 0.8$ & $54.2 \pm 0.8$ & \underline{$58.7 \pm 0.6$} & $18.9 \pm 0.4$ & $38.2 \pm 0.2$ & $42.6 \pm 0.5$ \\
                        & MobTcast  & \underline{$29.5 \pm 0.6$} & $51.3 \pm 0.7$ & $56.2 \pm 1.0$ & $17.3 \pm 0.6$ & $39.3 \pm 0.4$ & $43.4 \pm 0.9$ \\
                        & Ours (MHSA)     & $\bm{31.4 \pm 0.9}$ & $\bm{56.4 \pm 0.4}$ & $\bm{60.8 \pm 0.8}$ & \underline{$21.8 \pm 1.0$} & $\bm{42.5 \pm 0.7}$ & $\bm{46.5 \pm 0.3}$ \\ \bottomrule

    \end{tabular}

}
\end{table*}

We report that FPMC outperforms the 1-MMC method on all indicators except for the F1 score. The performance gap is large on Acc$@$5, Acc$@$10, MRR, and NDCG$@$10, implying that FPMC can better identify user preferences by considering collective mobility knowledge.
The relatively high F1 score of the 1-MMC method suggests that this method is practical if the focus is on the prediction performance of essential locations.
However, their performances are still significantly inferior to DL-based models on the considered datasets.
This difference emphasizes the importance of considering long-range dependencies and spatio-temporal contexts in the prediction task.

Compared with other DL baselines, the proposed MHSA-based network performs best for all indicators on both datasets.
For example, we report absolute increases of 0.8\% for the GC dataset and 1.6\% for the Geolife dataset on Acc$@$1 compared to the second-best performing method.
This performance gain results from its effective architectural design, including the FC residual block and multiple sets of self-attention layers, which explicitly capture long-term dependencies and periodicity in individuals' location visits.
We can also observe the importance of the attention module by comparing the LSTM and LSTM attn models.
With an additional self-attention layer between the hidden states, LSTM attn achieves higher scores on Acc$@$1, F1 score, MRR, and NDCG$@$10, and comparable performances on Acc$@$5 and Acc$@$10.
This result indicates that adding a single layer of attention in LSTM effectively identifies the most likely next location visit but does not increase the capability to extract the location choice set.
However, Deepmove and MobTcast, which both include the attention mechanism, do not achieve state-of-the-art results on our datasets.
As they are explicitly designed for check-in datasets, these models contain excessive components that might not help capture individuals' mobility transition, an essential property in mobility datasets with high temporal tracking coverage.
We present complementary experiments on check-in data in Appendix~\ref{app:check-in} to evaluate the generalizability of the next location prediction models in recommending the next POI.

\subsection{Influence of spatio-temporal context}

We perform an ablation study to understand the importance of spatio-temporal contexts in the MHSA-based model.
Table~\ref{tab:ablation} regards the model that includes time and user features as the baseline and reports the variation in the prediction performances when altering the combination of input features, including user, activity duration, and POI.
The ablation on the POI feature is only conducted for the GC dataset as we do not possess it in the Geolife dataset.
The performance variations and trends are consistent for both datasets.

%

\begin{table}[htbp!]
    \centering
    \normalsize

    \caption{Ablation study of the spatio-temporal context. The baseline model considers time and user features.}
    \label{tab:ablation}
    \makebox[\textwidth]{
        \begin{tabular}{@{}clcccccccc@{}}
        \toprule
        Dataset & Models & Acc@1 & Acc@5 & Acc@10 & F1 & MRR & NDCG@10    \\ \midrule
        
        \multirow{6}{*}{GC}      & Baseline                 & $42.5 \pm 0.1$ & $63.6 \pm 0.2$ & $68.6 \pm 0.1$ & $32.3 \pm 0.3$ & $52.1 \pm 0.1$ & $55.7 \pm 0.1$\\
                                 & Baseline $-$ User        & $42.3 \pm 0.2$ & $63.1 \pm 0.1$ & $68.1 \pm 0.2$ & $31.9 \pm 0.4$ & $51.8 \pm 0.1$ & $55.4 \pm 0.1$\\
                                 & Baseline $+$ sPOI        & $42.7 \pm 0.2$ & $64.2 \pm 0.1$ & \underline{$69.2 \pm 0.1$} & $32.3 \pm 0.2$ & $52.5 \pm 0.1$ & $56.1 \pm 0.1$\\
                                 & Baseline $+$ POI (TF-IDF)& $42.7 \pm 0.2$ & \underline{$64.3 \pm 0.2$} & \underline{$69.2 \pm 0.2$} & $32.4 \pm 0.1$ & $52.5 \pm 0.3$ & $56.2 \pm 0.1$ \\
                                 & Baseline $+$ POI         & $43.0 \pm 0.1$ & \underline{$64.3 \pm 0.1$} & \underline{$69.2 \pm 0.2$} & $32.4 \pm 0.1$ & $52.8 \pm 0.1$ & $56.2 \pm 0.1$\\
                                 & Baseline $+$ Duration    & \underline{$45.0 \pm 0.1$} & $64.0 \pm 0.1$ & $68.7 \pm 0.1$ & \underline{$34.0 \pm 0.3$} & \underline{$53.6 \pm 0.1$} & \underline{$56.9 \pm 0.1$}\\ 
                                 & All                      & $\bm{45.6 \pm 0.1}$ & $\bm{64.7 \pm 0.1}$ & $\bm{69.5 \pm 0.1}$ & $\bm{34.8 \pm 0.2}$ & $\bm{54.3 \pm 0.1}$ & \bm{$57.6 \pm 0.1$}\\ \midrule
        \multirow{3}{*}{Geolife} & Baseline                 & \underline{$29.8 \pm 0.9$} & \underline{$53.4 \pm 2.2$} & \underline{$58.3 \pm 2.1$} & $\bm{20.7 \pm 0.8}$ & \underline{$40.6 \pm 1.4$} & \underline{$44.9 \pm 0.8$}\\
                                 & Baseline $-$ User        & $28.3 \pm 1.0$ & $49.7 \pm 1.5$ & $54.5 \pm 0.7$ & $17.8 \pm 1.1$ & $38.2 \pm 0.9$ & $42.1 \pm 0.4$\\
                                 & Baseline $+$ Duration    & $\bm{31.4 \pm 0.9}$ & $\bm{56.4 \pm 0.4}$ & $\bm{60.8 \pm 0.8}$ & $\bm{21.8 \pm 1.0}$ & $\bm{42.5 \pm 0.7}$ & \bm{$46.5 \pm 0.3$}\\ \bottomrule
        \end{tabular}
    }
\end{table}

We observe that the various spatio-temporal contexts have distinct types of influence on the prediction performance.
The user feature improves the model's performance on all indicators.
This increase shows that merely processing the sequential information is insufficient to distinguish between users, and learning from explicit user information with the FC residual block helps identify personal preferences.
Moreover, activity duration and POI both contribute positively to the prediction performances.
While models that consider activity duration notably increase in Acc$@$1, F1 score, MRR, and NDCG$@$10, e.g., an absolute increase of 2.5\%, 1.7\%, 1.5\%, and 1.2\%, respectively, on GC, the POI feature has a more prominent effect on Acc$@$5, Acc$@$10, with an absolute increase of 0.7\% and 0.6\%, respectively on GC.
We argue that the activity duration is user-dependent information that describes personal preferences, whereas POI and its inferred functional land use are user-independent contexts that delineate location semantics.
Therefore, the former assists in identifying the correct next location, and the latter helps narrow down the possible location choice set in the prediction.
Regarding the encoding of POI context for locations, we first evaluate the performance of our model when using a single spatial scale ($dist = 250$ meters) to generate the land use context for each location (denoted as sPOI in Table~\ref{tab:ablation}).
sPOI encodes the type of POIs surrounding each location through BOW and LDA, but does not capture the distance information between POIs and locations. The evaluation results show a slight drop in performance across all indicators, except for Acc$@$10 and F1 score, compared to the multi-scale representation, indicating that land use context is better described at multiple spatial scales. 
Additionally, we conduct an experiment using the TF-IDF method that quantifies POI category intensities for locations (denoted as POI (TF-IDF)). To ensure compatibility, we introduce an extra learnable linear layer that maps the number of POI categories to the number of topics \textit{tp} in LDA before applying the context network $f^{\boldsymbol{C}}(\cdot; \cdot)$. POI (TF-IDF) yields slightly worse results in Acc$@$1 and MRR than the proposed model that uses LDA.
Overall, the models that include all contexts achieve the best performance, with an Acc$@$1 increase of 3.1\% in GC and 1.6\% in Geolife compared to the baseline model, suggesting that these contexts represent different aspects of individual mobility and all contribute to an accurate next location prediction result.

\subsection{Individual vs.\ collective models}

\begin{figure}[htbp!]
  \centering
  \begin{subfigure}[]{0.4\textwidth}
    \centering
    \includegraphics[height=3.4cm]{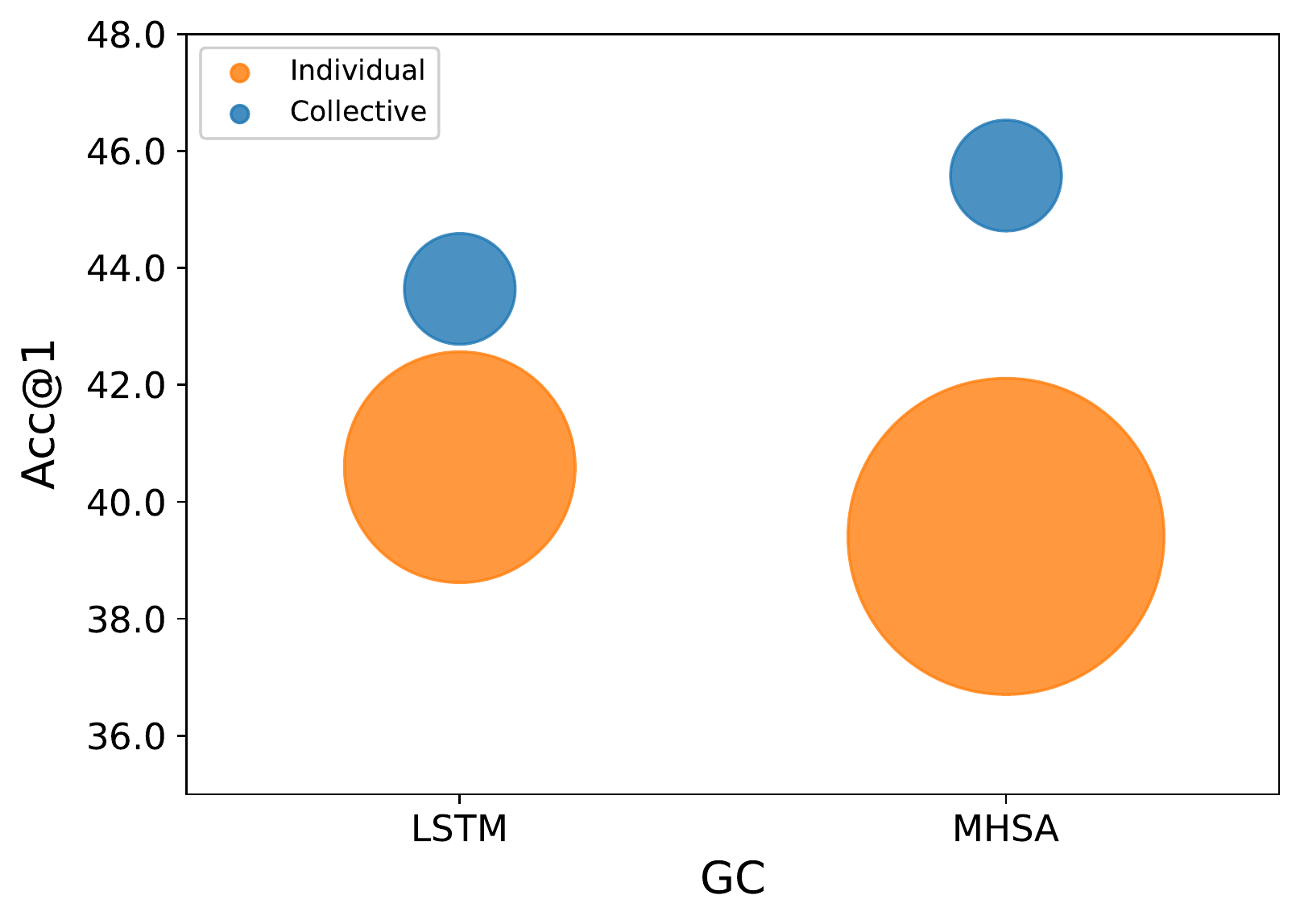}
  \end{subfigure}
  \begin{subfigure}[]{0.4\textwidth}
    \centering
    \includegraphics[height=3.4cm]{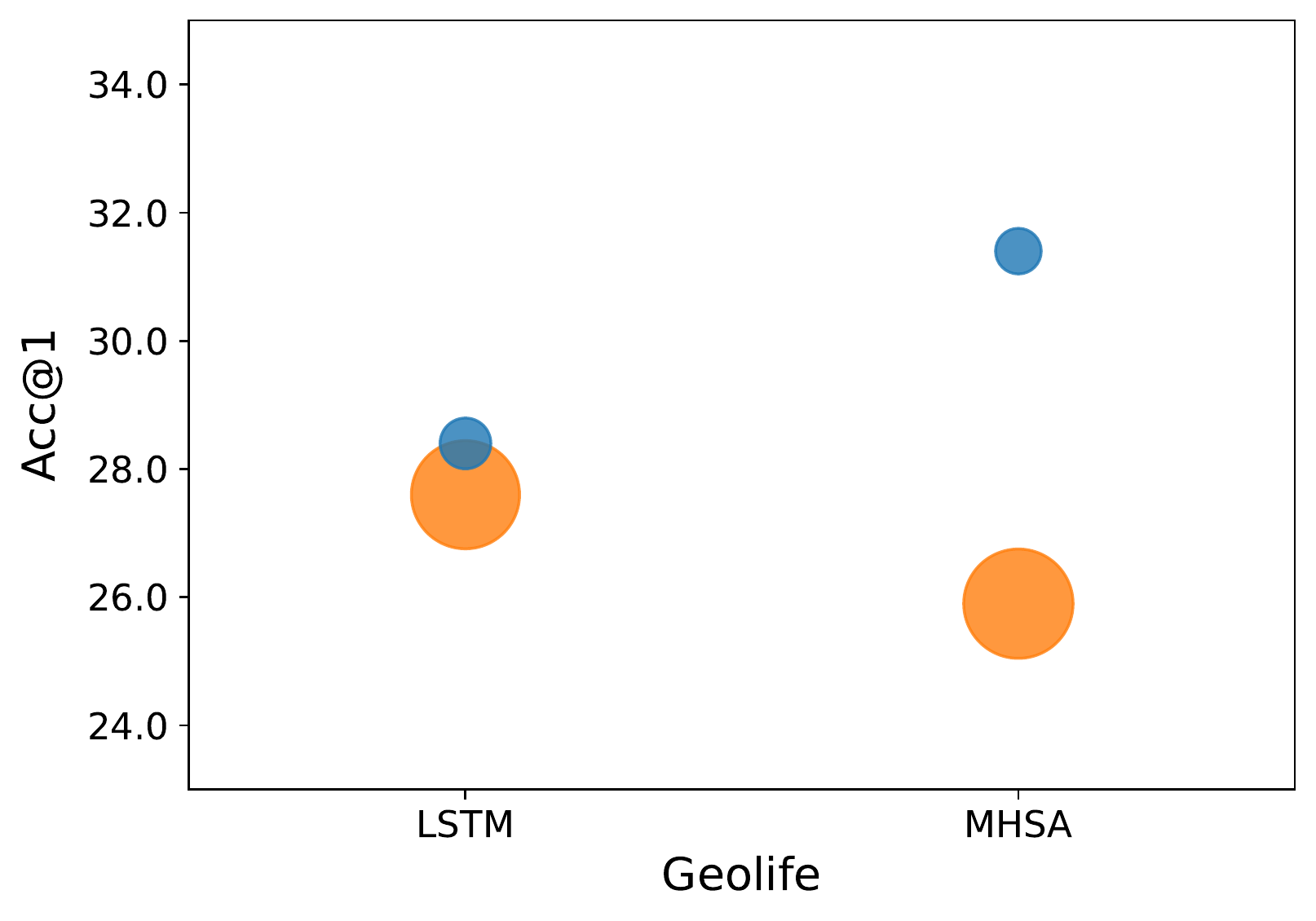}
  \end{subfigure}%
  \begin{subfigure}[]{0.18\textwidth}
    \centering
    \includegraphics[height=3cm]{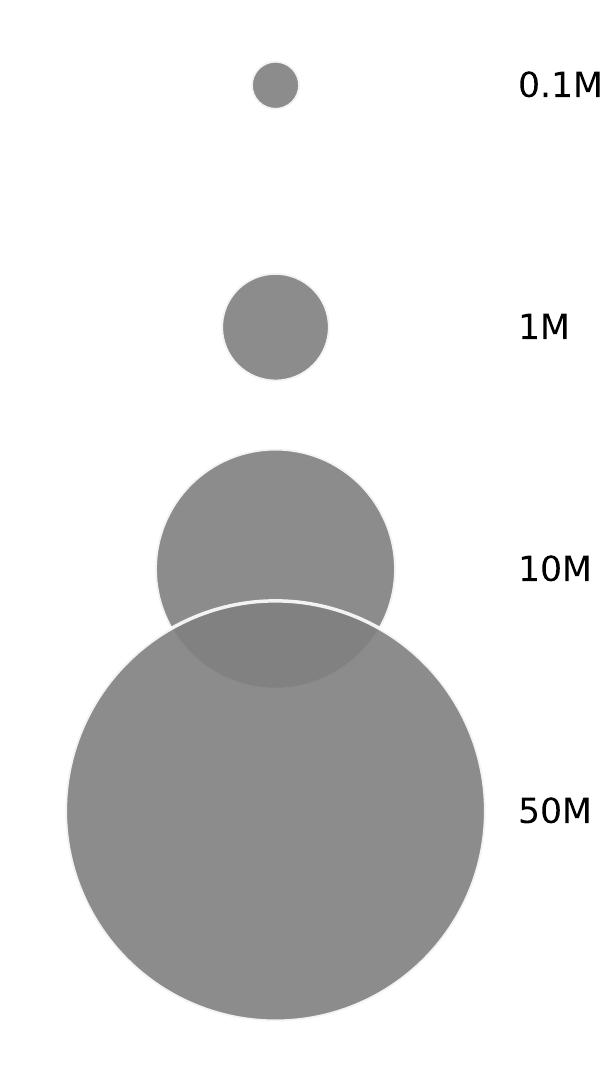}
  \end{subfigure}%

  \caption{Individual vs. collective models' performance. We consider the Acc$@$1 score of the LSTM and the MHSA-based model for the GC and Geolife datasets. The diameter size denotes the model's total number of parameters (M: million).}
  \label{fig:granularity}
\end{figure}

To study the granularity choice, we reveal the key differences between training a model on the whole population dataset and maintaining personalized models for each user.
We choose LSTM and the proposed MHSA-based model for the comparison in this section.
Figure~\ref{fig:granularity} shows their Acc$@$1 scores for models trained for each individual (individual) and on the whole population (collective), with the diameter size illustrating the total number of parameters for the respective model. More details about the comparison can be found in Appendix~\ref{app:compare}.

We find that collective models consistently achieve a higher prediction accuracy with fewer model parameters, showing models trained on population data can infer certain collective mobility patterns beneficial to predicting a user's next location.
The performance increase is more prominent for the MHSA-based model than LSTM.
Interestingly, although the MHSA-based model outperforms LSTM on the collective level, the latter achieves a higher prediction accuracy when trained on individual-level data.
In the context of location prediction, this finding implies that the MHSA-based model has a stronger capability to learn complex dependencies when abundant data are provided, whereas LSTM is more efficient with fewer training instances.
Furthermore, considering user information in the collective model benefits the prediction performance, as shown from the ablation results in Table~\ref{tab:ablation}. This phenomenon reveals that a balance between collective patterns and individual preferences takes effect in mobility prediction models.
Overall, we provide evidence that collective-level models should still be preferred for next location prediction on continuously recorded mobility traces.

\begin{figure}[htbp!]
  \centering
  \includegraphics[width=0.7\linewidth]{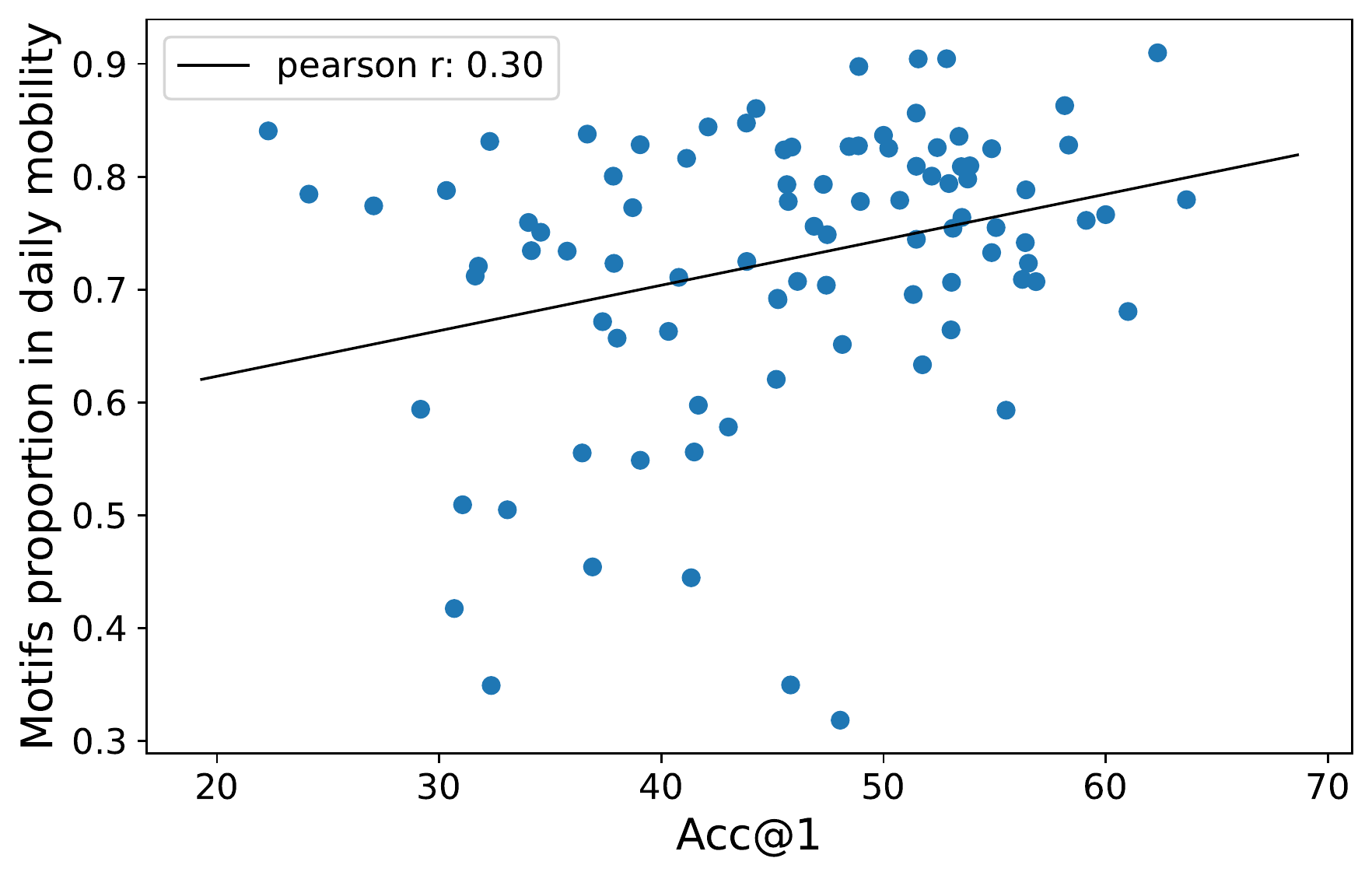}
  \caption{Relation between the prediction accuracy of the MHSA-based network and the proportion of motifs in daily mobility. The motif proportion of a user is positively correlated with the accuracy obtained from the network.}
  \label{fig:motifs}
\end{figure}

To understand the types of collective patterns captured by the DL network, we employ the concept of mobility motif, which quantifies the regularity of a user's daily mobility.
Mobility motif refers to the recurring network patterns that emerge when individual daily location visits are abstracted as networks, with nodes representing activity locations, and links representing location transitions~\citep{schneider_unravelling_2013}.
Here, we use the GC dataset, where the tracking quality of users is high (see Appendix~\ref{app:data}), and regard individual daily location networks that appear on average more often than 0.5\% in the dataset as mobility motifs. As a result, we identify 16 motifs that can capture, on average, 73\% of the daily location transition patterns.
These motifs are shared among users, indicating the presence of common mobility patterns. The proportion of these motifs in a user's mobility provides insights into the prevalence of these shared patterns. 
Consequently, we anticipate a positive relationship between the proportion of motifs and the performance of the next location prediction model, as it suggests that the network is effectively learning these shared mobility patterns.
This relation is illustrated in Figure~\ref{fig:motifs}, where we observe a weak positive correlation between the motifs proportion and Acc$@$1 across users (Pearson correlation coefficient $\rho=0.30$, two-tailed $P <0.01$).
This result indicates that users with a higher prevalence of repeated mobility patterns also achieve higher location prediction accuracy from the DL network.
Although motifs merely provide a high-level description of individual mobility, the correlation provides evidence that the prediction network successfully captures collective mobility patterns.

\subsection{Impact of historical input lengths}

\begin{figure}[htbp!]
  \centering
  \includegraphics[width=0.7\linewidth]{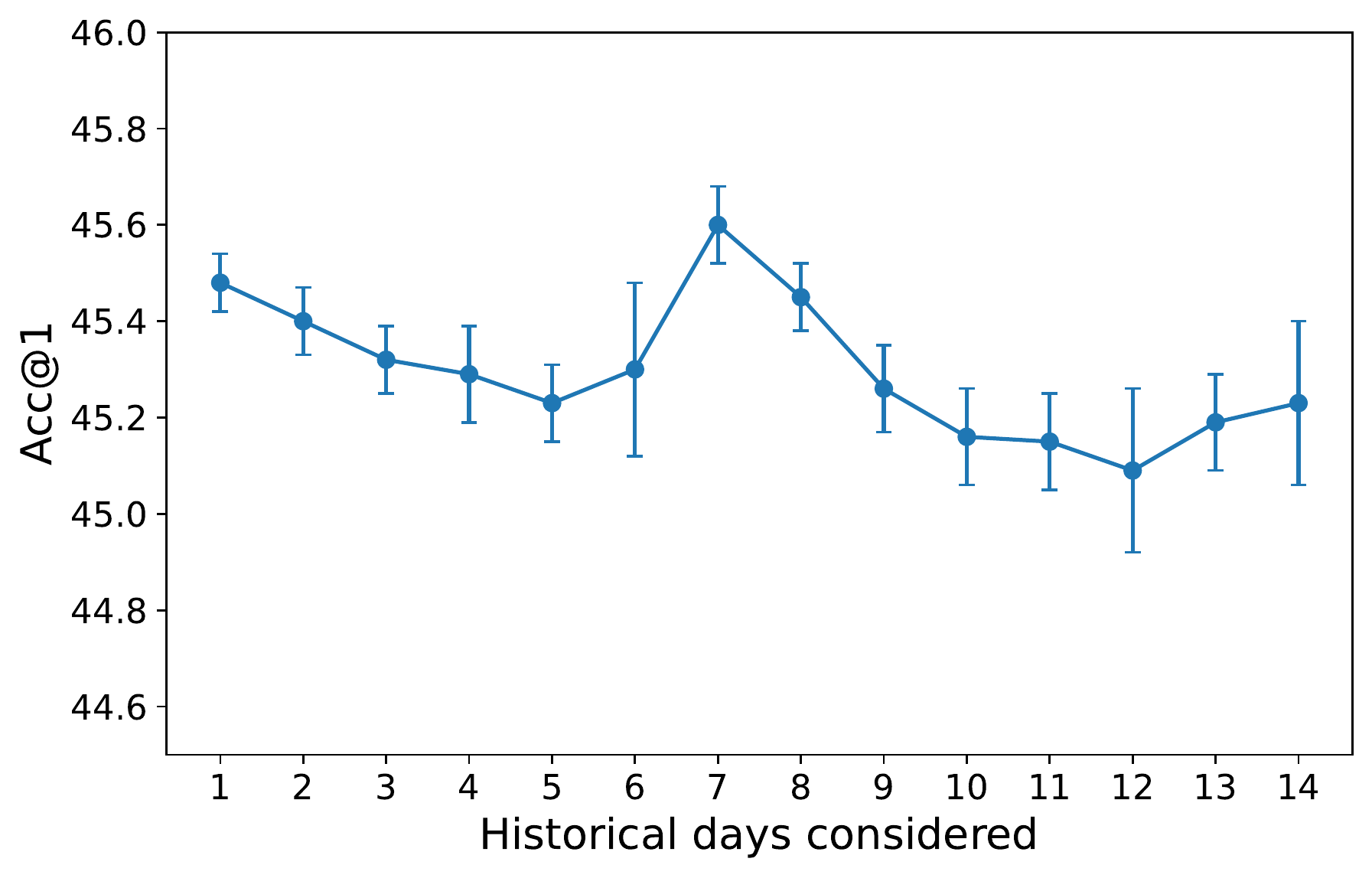}
  \caption{Line plot showing the change in prediction accuracy through altering the length of the input sequences. Error bars show the standard deviation across five runs. The network that accepts sequences visited in the past 7 days obtains the highest accuracy.}
  \label{fig:length}
\end{figure}

Next, we are interested in how much historical knowledge should be considered for the DL network to achieve the desired performance.
We identify the visit time of each historical record with reference to the current prediction and alter the length of the input sequence by controlling the number of days $D$ to consider in the past.
Figure~\ref{fig:length} illustrates the Acc$@$1 of the proposed MHSA model when changing the input sequence length.
An overall decreasing trend can be observed when the number of considered historical days increases, meaning that including longer sequences, and thus more information, does not necessarily lead to better model performance.
Moreover, we observe two peaks in the Acc$@$1 trend corresponding to stay point visits in the previous 7 and 14 days.
The results of Mann-Whitney U tests show that the Acc$@$1 obtained from the past 7 days is not significantly different than considering 1 day in the past (two-tailed $P =0.09$), but is significantly different than the one obtained from all other input lengths (all two-tailed $P <0.05$).
Therefore, we conclude that the proposed MHSA-based model achieves the best performance when considering mobility conducted in the past 7 days.
Also, the performance peaks suggest that mobility traces from one and two weeks ago carry additional information that is beneficial to predicting the current day's location visit.

\begin{figure}[htbp!]
  \centering
  \begin{subfigure}[]{0.7\textwidth}
    \centering
    \includegraphics[width=\textwidth]{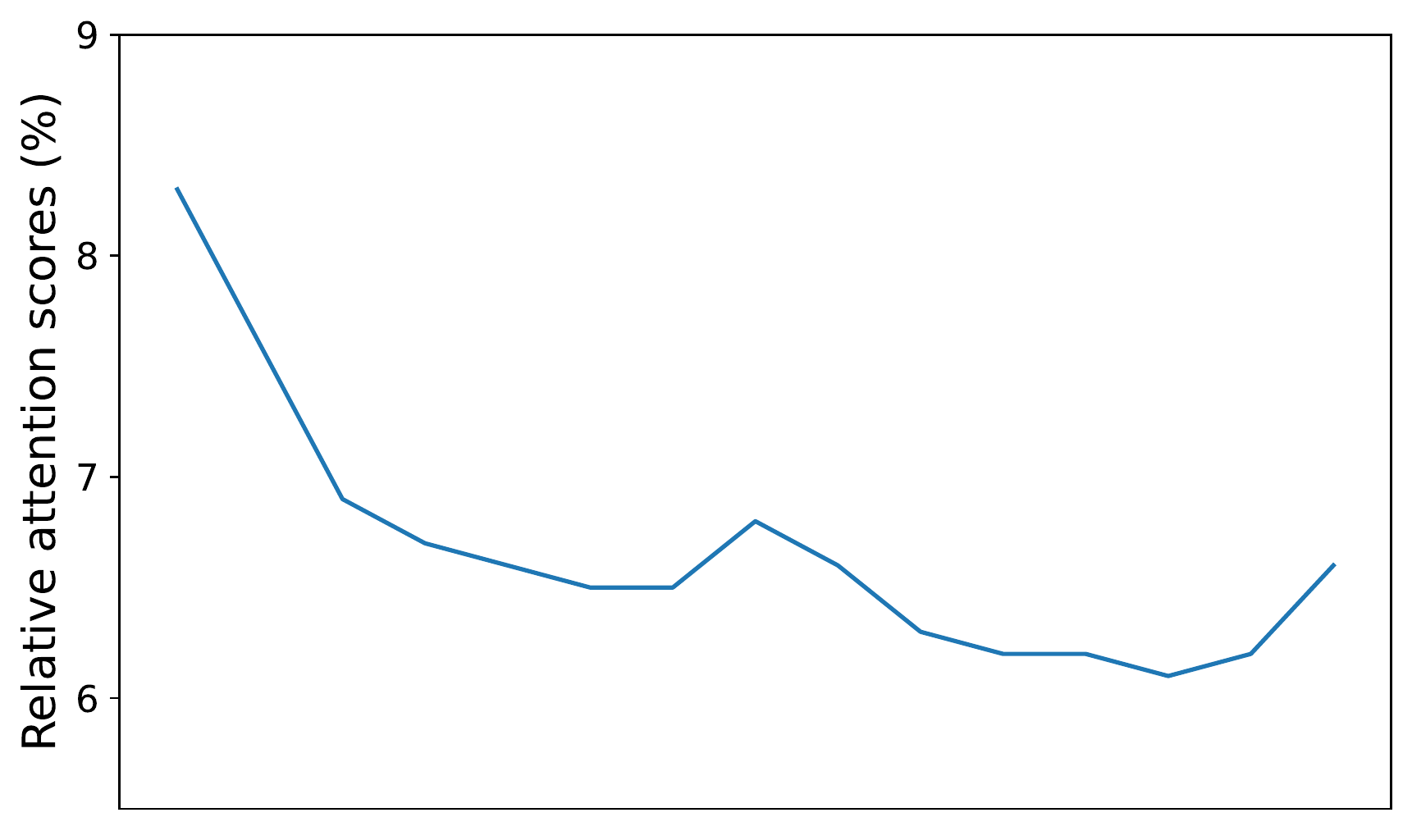}
  \end{subfigure}
  \begin{subfigure}[]{0.7\textwidth}
    \centering
    \includegraphics[width=0.93\textwidth, right]{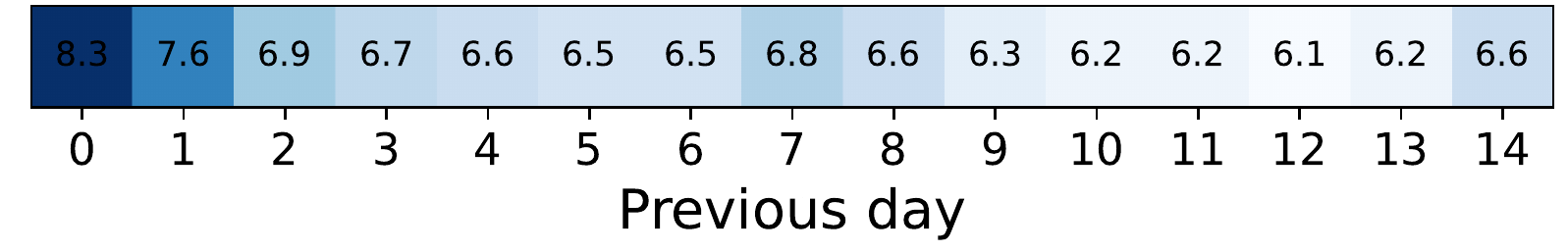}
  \end{subfigure}%

  \caption{Relative attention weights of different previous days considered by a trained network. The stay points visited on day 0 (current day), 1 (previous day), 2 (two days before), 7 (one week before), and 14 (two weeks before) received relatively
  high attention. }
  \label{fig:transformer}
\end{figure}

To reveal a prediction's temporal dependence on the input sequence, we visualize the attention weights within the proposed MHSA model.
The MHSA model learns the sequential relation of input records and their relative importance directly from data.
These learned dependencies could be reflected in the attention weights of each time step, with a larger weight indicating a higher contribution to the network's prediction.
We extract the attention weights of test dataset predictions based on a model trained using location visits in the past 14 days.
Then, we calculate the average attention weights of all network layers and visualize the average weights according to their corresponding previous days, as shown in Figure~\ref{fig:transformer}.
We observe that the trained network focuses on stay points travelled in recent days, such as the current day and the day before, which intuitively suggests the short-term interdependence of mobility activities.
Moreover, the records from 7 and 14 days ago receive relatively high attention, further verifying the finding in Figure~\ref{fig:length}.
Together with the performance results, we conclude that weekly periodicity is important for mobility prediction.

We finally identify the essential mobility records, with which the trained model achieves the best prediction result.
As individuals typically start and end their daily mobility at the same location (e.g., home), separating individual mobility traces according to natural days will barely lose any location transition information.
Therefore, we divide input stay points into sub-sequences according to their recorded day and construct new input sequences by selecting different combinations of these sub-sequences.
The ablation study on input sequences in Table~\ref{tab:day_ablation}, where we denote models trained with different combinations of input records, shows distinct performances of the MHSA-based model when altering the considered information.
The records on the current day are indispensable for predicting the next location visit (V1 vs. Figure~\ref{fig:length}). Although including more information from the past days slightly improves the prediction performance (V1 vs. V2), the model's prediction ability is still far behind other model variations.
Moreover, we report the necessity of considering the previous day (day 1), as shown from the performance gain when comparing model variations V3 and V5, as well as V4 and V6.
However, we observe a drop in performance when including records from 2 days before (V5 vs.\ V7) and from 14 days before (V3 vs.\ V4 and V5 vs.\ V6), suggesting that this information might be repetitive and less critical for the current prediction.
In short, this ablation study shows that instead of feeding the whole historical sequence into the model, selecting only the records from the current day, the previous day, and the day one week before is sufficient to achieve optimum performance for next location prediction.

\begin{table}[htbp!]
\centering
\caption{Ablation study of historical dependency. We bin the stay points according to the time of visit w.r.t. the current time step, such that 0 represents the visits on the same day, and 14 includes the ones visited two weeks before. We report the network's performance on a subset of input sequences, where only records on certain previous days are included.}
\label{tab:day_ablation}
\begin{tabular}{@{\extracolsep{\fill}}cccccccccc@{}}\toprule
                         & \multicolumn{5}{c}{Previous days}                                                                                         &                         & \multicolumn{1}{c}{}                      & \multicolumn{1}{c}{}                     \\ \cmidrule(lr){2-6}
\multirow{-2}{*}{Models} & 0                        & 1                        & 2                        & 7                        & 14                       & \multirow{-2}{*}{Acc@1} & \multicolumn{1}{c}{\multirow{-2}{*}{F1}} & \multicolumn{1}{c}{\multirow{-2}{*}{MRR}} & \multicolumn{1}{c}{\multirow{-2}{*}{NDCG@10}}  \\ \midrule

V1                       & -          & \checkmark & -          & -          & -          & $35.9 \pm 0.2$ & $23.9 \pm 0.1$ & $48.6 \pm 0.2$ & $52.9 \pm 0.1$                                       \\
V2                       & -          & \checkmark & \checkmark & \checkmark & \checkmark & $36.6 \pm 0.2$ & $24.6 \pm 0.3$ & $49.0 \pm 0.1$ & $53.3 \pm 0.1$        \\                        
V3                       & \checkmark & -          & -          & \checkmark & -          & $44.9 \pm 0.2$ & $34.2 \pm 0.3$ & $53.8 \pm 0.1$ & $57.2 \pm 0.1$                                       \\
V4                       & \checkmark & -          & -          & \checkmark & \checkmark & $44.6 \pm 0.2$ & $34.0 \pm 0.2$ & $53.6 \pm 0.2$ & $56.9 \pm 0.2$                                       \\ 
V5                       & \checkmark & \checkmark & -          & \checkmark & -          & $\bm{45.6 \pm 0.2}$ & $\bm{34.9 \pm 0.2}$ & $\bm{54.3 \pm 0.1}$ & $\bm{57.6 \pm 0.2}$                                        \\
V6                       & \checkmark & \checkmark & -          & \checkmark & \checkmark & $\bm{45.5 \pm 0.2}$ & $\bm{35.0 \pm 0.2}$ & $\bm{54.3 \pm 0.1}$ & $\bm{57.5 \pm 0.2}$                                   \\ 
V7                       & \checkmark & \checkmark & \checkmark & \checkmark & -          & $\bm{45.5 \pm 0.1}$ & $\bm{34.8 \pm 0.4}$ & $\bm{54.3 \pm 0.1}$ & $\bm{57.6 \pm 0.1}$                                          \\\bottomrule

\end{tabular}
\end{table}

\section{Discussion and conclusion}\label{sec:discussion}
We have introduced a deep learning model that uses MHSA for next location prediction.
The model is designed to learn transition patterns from historical location visits, their visit times and activity duration, as well as their surrounding land use functions, to infer accurate location prediction results for each user.
In particular, we propose representing the locations' land use context at multiple spatial scales with POI data and LDA.
Through experiments with two large-scale longitudinal GNSS tracking datasets, we demonstrate the effectiveness of our proposed model in comparison with classical and state-of-the-art learning-based models.
We conduct an ablation study to investigate the importance of spatio-temporal context in the next location prediction task, aiming to enhance the interpretability of the model.
We conclude that distinguishing sequences recorded from different users is indispensable for the model to learn personal preferences. We observe that including the user-dependent feature (activity duration) increases the performance of predicting the exact next location, and considering user-independent context (functional land use) helps reduce the size of the possible location choice set.
These results reveal the distinct influence of various features on prediction performances, help identify the most relevant contexts for specific tasks, and enhance our understanding of mobility context modelling.

Based on the state-of-the-art MHSA model, we discuss different modelling choices for the next location prediction.
We find that models trained with population data achieve higher prediction performance with fewer parameters than individual-level models, suggesting the collective-level network captures common movement patterns.
Using the mobility motif as a proxy for shared mobility patterns, we find a correlation between users' mobility regularity and the network's prediction performance.
Combining these findings with the results from the ablation study, we conclude that mobility prediction models should strive for a balance between collective patterns and personal mobility preferences - we recommend training individual prediction models on collective-level data and incorporating personal information through the user identifier.
However, individual-level models may still be valuable when sample sizes are too small to summarize common movement patterns from the population.
Moreover, we report that the model trained with the past seven days of records achieves the highest prediction performance compared to all other historical sequence lengths.
Visualizing attention weights within the MHSA model shows apparent daily and weekly dependencies.
However, not all historical records are equally important for accurate prediction. The model achieves an optimum location prediction result by learning from the records from the current day, the previous day, and the day one week before.
This analysis enhances the interpretability of the model regarding historical dependencies and offers insights into the significance of periodic patterns in predicting individual mobility. Moreover, it allows for more efficient network training and potentially saves data storage, as prediction models do not need to access the entire movement history.

We believe the proposed MHSA-based next location prediction network is an important step towards accurate and interpretable individual mobility prediction, promoting the implementation of downstream applications. The model can be readily applied to populations engaged in mobility-as-a-service or other personalized mobility services, and serve as a backbone for planning on-demand transport services~\citep{kieu_class_2020}, implementing mobility incentives~\citep{xiong_integrated_2020}, and suggesting alternative mobility options~\citep{bucher_location_2019}.

We see several future directions based on the results of this study.
1) We focused on predicting the next location for users with an accessible location history. As a follow-up study, the proposed framework can be generalized for new or previously unknown users, which will be valuable for large-scale policy design and evaluation. This generalization requires sufficient learning of collective mobility patterns through training on datasets with a large and representative population, as well as designing components in the network to measure similarities across users explicitly. Representation learning approaches that describe locations from physical and socio-economic perspectives~\citep{bai_geographic_2023} can be integrated into the framework to enhance the characterisation of less-visited locations. 
2) Additionally, the prediction performance of the network may be affected by user-specific factors, such as the length and quality of the historical tracking information. Therefore, future studies should aim to quantify these dependencies to gain a better understanding of personalized location recommendations.
3) We use LDA vectors to represent land use functions surrounding locations. This approach can be extended by incorporating more recent learning-based POI embedding methods, such as Word2Vec POI embedding~\citep{yao2017sensing} and the Place2Vec model~\citep{zhai2019beyond}, which capture the spatial distribution of POIs within each study unit. Quantifying the benefits of these approaches for location prediction and determining the optimal representation of POIs are exciting directions for future research.
4) Lastly, the relation between spatio-temporal contexts and mobility is more evident on a longer time scale than the choice of the next activity location. Future studies should investigate context integration for long-term mobility prediction and generation tasks.
Overall, we anticipate that this study will raise attention to spatio-temporal context integration for predicting human mobility, as well as help design and implement mobility prediction models.

\bibliography{mybibfile}

\beginsupplement

\newpage
\section*{Appendix}

\subsection{Movement definition}\label{app:movement}
\begin{figure}[htbp!]
  \centering
  \includegraphics[width=0.6\linewidth]{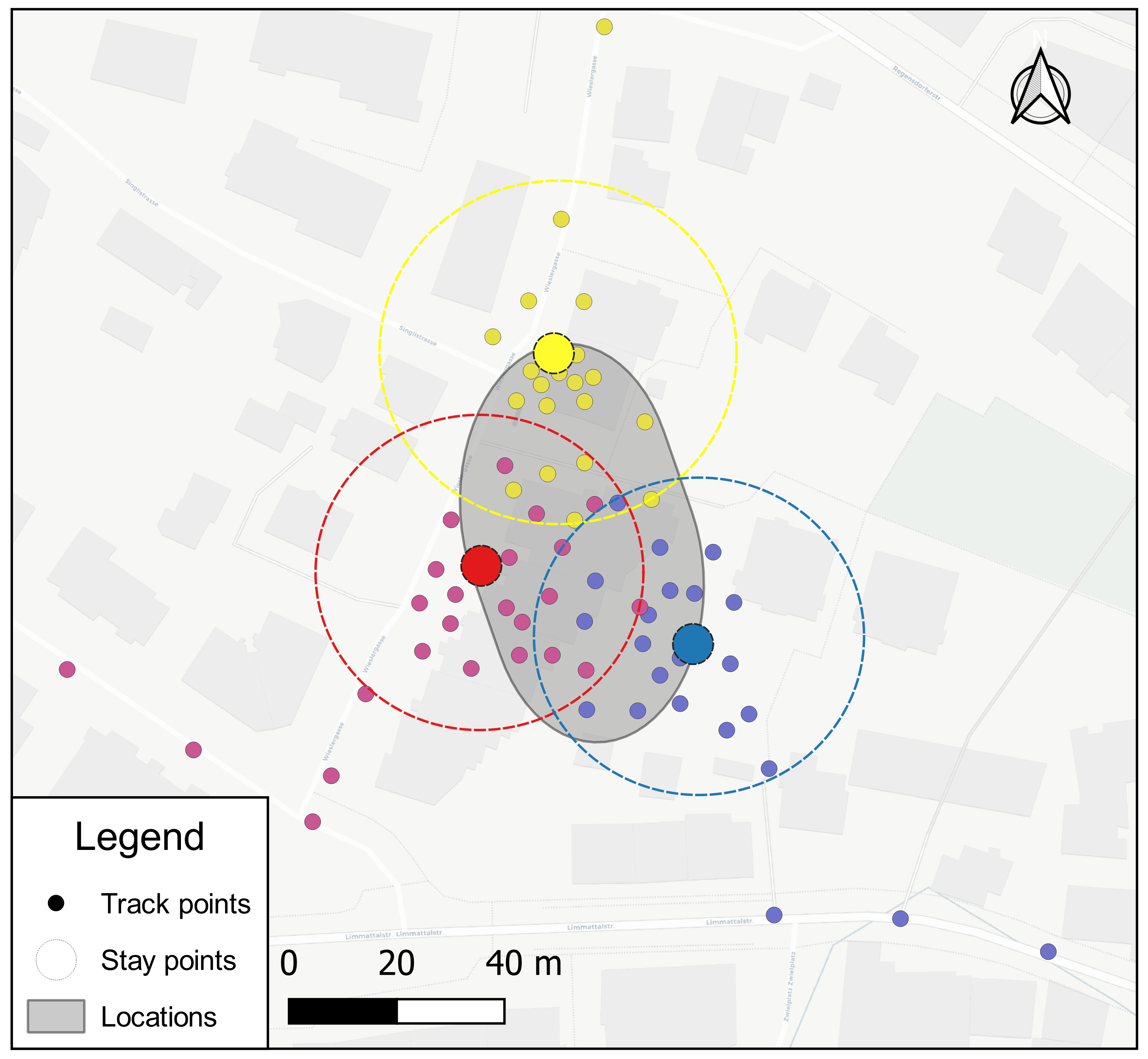}
  \caption{Graphic example of the movement data generation process. Stay points and locations are generated through spatially aggregating GNSS track points. Mobility traces were recorded by one of the authors living in Zurich, Switzerland. Map data ©OpenStreetMap contributors, ©CARTO.}
  \label{figAppend:locations}
\end{figure}

Figure~\ref{figAppend:locations} depicts the process of generating stay points and locations from GNSS track points. This process attaches different semantic levels to movement data, which is useful for transportation and mobility applications~\citep{bucher_location_2019}. In addition, it retains the most critical activity information for next location prediction while significantly reducing the amount of data that must be analysed. We refer to~\citet{axhausen_definition_2007} and ~\citet{Martin_trackintel_2023} for the conceptual definition and the implementation of the movement data model. 

\subsection{Network and implementation}\label{app:network}
\paragraph{\textbf{Network architecture}}

\begin{figure}[htb!]
  \centering
  \includegraphics[width=1\linewidth]{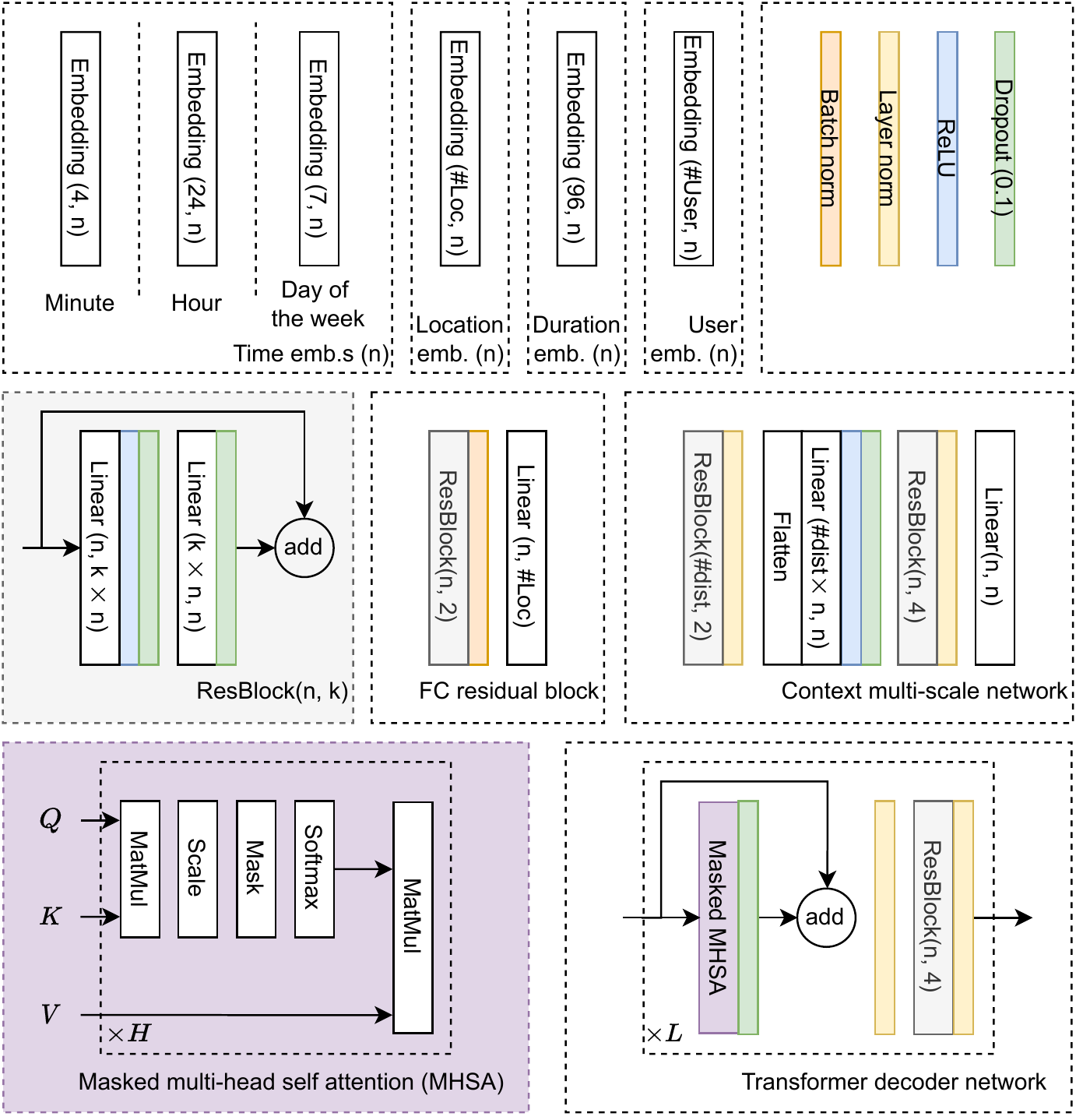}
  \caption{Detailed network architecture. $(n, m)$ in embedding and linear layers denote the input and output feature dimension. $(n, k)$ in the residual block denotes the input feature dimension and the feature scaling within the block.}
  \label{figAppend:architecture}
\end{figure}

The detailed embedding layers and network architecture is depicted in Figure~\ref{figAppend:architecture}. We introduce separate embedding layers for the location identifier, the visit time, and the activity duration of each stay point in the historical sequence, as well as the user identifier from whom the sequence was recorded. The ResBlock that consists of linear layers with residual connection~\citep{He_16} is the backbone of the FC residual block, the context multi-scale network, and the Transformer decoder network.

\paragraph{\textbf{Multi-head self-attention}}
The multi-head self-attention mechanism is one indispensable component that supports the effectiveness of the transformer architecture. The attention operation is commonly framed as obtaining an output value from a query and a set of key-value pairs. Query, key, and value are all vectors with the same size, which we denote as $dim$. Here, scaled dot-product attention is applied and efficiently implemented by packing a set of query, key, and value vectors into matrices Q, K, and V~\citep{Vaswani_2017}:
\begin{linenomath}
  \begin{equation}
    \text{Attention(}Q, K, V\text{)} = \text{softmax(}\frac{QK^T}{\sqrt{dim}})V
  \end{equation}
\end{linenomath}

Then, multi-head attention is constructed by scaling the matrices Q, K, and V through parameter matrices and concatenating the results of $H$ attention functions:
\begin{linenomath}
  \begin{align}
    \text{MultiHead(}Q, K, V \text{)} & = \text{cat[}head_1, ..., head_H]\boldsymbol{W}^O                                  \\
    \text{where} \quad head_i         & = \text{Attention(}Q\boldsymbol{W}^Q_i, K\boldsymbol{W}^K_i, V\boldsymbol{W}^V_i )
  \end{align}
\end{linenomath}
where $\text{cat}[\cdot]$ denotes concatenation. Within the network, key, value, and query matrices are identical and correspond to the output of the previous block. In the first block, they are set as the input to the network, i.e., the embedding matrix $\boldsymbol{e}^{all}$, obtained through stacking overall embedding vectors along their sequence order. In addition, forward-masking operations are included in the attention operations to prevent them from accessing information from ``future'' time steps; that is, the entry at step $k$ can only focus on the information preceding (and including) $k$.

\paragraph{\textbf{Implementation and training details}} A batch size of 256 and 32 is used for the GC and Geolife datasets, respectively. To ensure a consistent sequence length within each batch, any shorter sequences are padded with zeros at the end until they match the length of the longest sequence. The number of topic $tp$ in the LDA analysis is set to 16. The final selected network configurations are shown in Table~\ref{tab:conf}, determined through a grid search based on the performance of the network on the validation set:

\begin{sloppypar}
  GC dataset: the number of layers $L$ from $\{2, 4, 6\}$, the number of heads $H$ from $\{2, 4, 8\}$, the size of the overall embedding vector from $\{32, 64, 96, 128\}$, the dimension of the feedforward layers in the transformer network from $\{128, 256, 512\}$ and the dropout of the FC residual block from $\{0.1, 0.2, 0.5\}$.
\end{sloppypar}

Geolife dataset: the number of layers $L$ from $\{2, 4, 6\}$, the number of heads $H$ from $\{2, 4, 8\}$, the size of the overall embedding vector from $\{16, 32, 64\}$, the dimension of the feedforward layers in the transformer network from $\{64, 128, 256\}$ and the dropout of the FC residual block from $\{0.1, 0.2\}$.

\begin{table}[htbp!]
  \caption{The selected best-performing hyper-parameter set for the considered datasets. }
  \label{tab:conf}
  \centering
  \begin{tabular}{@{}ccc@{}}
    \toprule
                     & GC  & Geolife \\ \midrule
    \#Layers L       & 4   & 2       \\
    \#Heads H        & 8   & 8       \\
    Embedding dim.   & 96  & 32      \\
    Feedforward dim. & 256 & 128     \\
    Dropout          & 0.1 & 0.2     \\ \bottomrule
  \end{tabular}
\end{table}

\paragraph{\textbf{Model parameters and computation times}} 

We compare prediction models' parameter number and computation time in Table~\ref{tab:parameter}. The number of parameters for 1-MMC and FPMC represents the total elements in the transition matrix and the parameter matrices, respectively. 1-MMC and FPMC are implemented in Python, and the DL models are implemented using PyTorch. The benchmarking is performed on a machine equipped with a single RTX 3090 GPU and a 16-core CPU.

\begin{table*}[htbp!]
\centering
\caption{Parameter number, training time, and inference time of prediction models.}
\label{tab:parameter}

    \makebox[\textwidth]{
    \begin{tabular}{@{}ccccccc@{}}
    \toprule
    \multirow{2}{*}{Method} & \multicolumn{3}{c}{GC}     & \multicolumn{3}{c}{Geolife} \\ \cmidrule(lr){2-4}\cmidrule(lr){5-7}
                        & \#Params (K) & Train (s) & Inference ($\mu$s) & \#Params (K) & Train (s) & Inference ($\mu$s) \\ \midrule
                        
    1-MMC       & 1,405 & 7 & $374 \pm 44$ & 174 & 1 &  $361 \pm 57$ \\
    FPMC        & 2,374 & $5893 \pm 312$ & $2376 \pm 1146$ & 115 & $637 \pm 8$ & $78 \pm 18$\\
    LSTM        & 1,995 & $315 \pm 68$ & $10 \pm 2$ & 147 & $74 \pm 7$ & $7 \pm 2$\\
    LSTM attn   & 2,094 & $268 \pm 15$ & $82 \pm 20$ & 172 & $94 \pm 9$ & $55 \pm 5$\\
    Deepmove    & 2,152 & $363 \pm 9$ & $91 \pm 22$ & 184 & $114 \pm 7$ & $60 \pm 4$\\
    MobTcast    & 3,817 & $314 \pm 11$ & $33 \pm 8$ & 170 & $157 \pm 13$ & $19 \pm 5$\\
    Ours (MHSA) & 2,000 & $342 \pm 22$ & $32 \pm 7$ & 112 & $132 \pm 15$ & $12 \pm 4$\\ \bottomrule

    \end{tabular}

}
\end{table*}

\subsection{Dataset analysis}\label{app:data}
\begin{figure}[htbp!]
  \centering
  \includegraphics[width=1.0\linewidth]{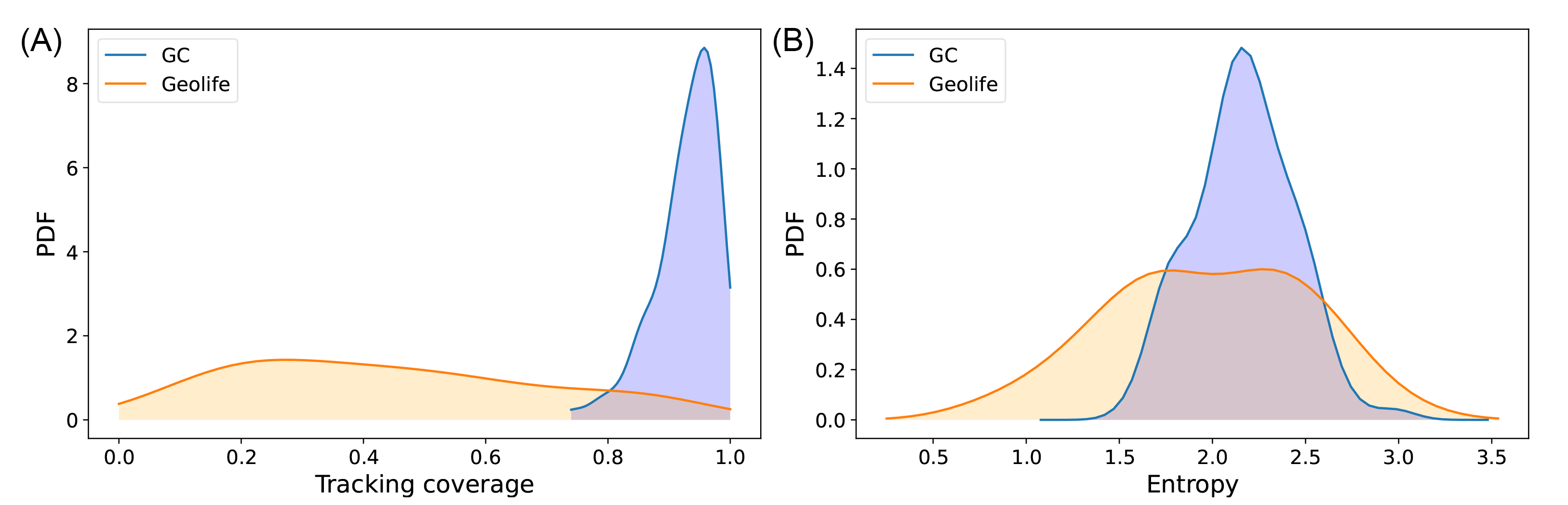}
  \caption{Dataset properties. (A) The overall temporal tracking coverage. (B) The entropy of the individual visited location sequences.}
  \label{fig:data}
\end{figure}

\paragraph{\textbf{Tracking quality}}
We utilize temporal tracking coverage to measure data quality in the temporal dimension~\citep{hong_conserved_2023}. Figure~\ref{fig:data}A shows the tracking coverage distribution of individuals for both considered datasets. We find that most individuals in the GC dataset have high tracking coverage ($>0.8$), whereas the coverage of the majority of Geolife users is low. Low tracking coverage indicates that we often lose track of the user's whereabouts during their daily mobility. The prevalence of these temporal gaps negatively influences next location prediction, as visits to locations might be missing, and incorrect location transition patterns are recorded.

\paragraph{\textbf{Mobility entropy}}
Mobility entropy captures the frequency of location visits and their visitation order~\citep{song_limits_2010}. It guarantees an upper bound for the theoretical predictability of individual location sequences and is often applied to compare mobility patterns across datasets~\citep{zhang_beyond_2022}.
Figure~\ref{fig:data}B shows the entropy distribution across users. We observe a similar sample average for both datasets. Still, the entropy distribution of Geolife is more dispersed than the one from GC, suggesting users with high theoretical predictability (low entropy) and low predictability (high entropy) coexist in the Geolife dataset. This could result from the wide range of temporal tracking coverage, but it also might be because the dataset contains users with distinct mobility behaviours.

\subsection{Complementary experiments on check-in data}\label{app:check-in}
To assess the generalizability of the location prediction models on check-in sequences, we conduct supplementary experiments using the publicly available Gowalla dataset~\citep{cho_friendship_2011} and Foursquare New York City (NYC) dataset~\citep{yang_modeling_2015}.
The Gowalla dataset comprises $\sim\!6.4$M check-ins from $\sim\!107$K users across multiple cities, while the Foursquare NYC dataset consists of $\sim\!227$K records from $\sim\!1$K users within a single city.
We follow common preprocessing standards for check-in data, which involve excluding unpopular POIs with fewer than 10 check-in records and filtering out users with fewer than 10 check-in histories~\citep{feng_deepmove_2018, sun_tcsa_2022}. 
After preprocessing, the location prediction performances of the implemented models are presented in Table~\ref{tab:check-in}.
Note that the MobTcast implementation does not incorporate users' social contexts and POIs' semantic category labels. 
Additionally, the proposed MHSA model does not consider land-use context and spent duration features for the check-in POIs.
In both datasets, we observe significant improvements with the LSTM attn compared to the baseline LSTM, and the MHSA model demonstrates slightly better performance than Deepmove and MobTcast.
These results highlight the effectiveness of attention-based methods and the proposed spatio-temporal embedding pipeline in uncovering sequential patterns of POI visits.
To summarize, we quantitatively evaluate the performance of location prediction models on check-in data, providing benchmarks for future reference.

\addtolength{\tabcolsep}{-2pt}
\begin{table*}[htb!]
\centering
\caption{Performance evaluation results for next location prediction on check-in datasets. }
\label{tab:check-in}

    \makebox[\textwidth]{
    \begin{tabular}{@{}cccccccc@{}}
    \toprule
    Dataset & Method    &  Acc@1 & Acc@5 & Acc@10 & F1  & MRR & NDCG@10      \\ \midrule
    \multirow{6}{*}{Foursquare NYC} & 1-MMC     & 16.0 & 32.5 & 36.4 & 14.4 & 23.1 & 26.6 \\
                        & FPMC      & $14.5 \pm 0.1$ & $40.4 \pm 0.4$ & $51.9 \pm 0.2$ & $7.0 \pm 0.1$ & $26.5 \pm 0.2$ & $31.8 \pm 0.2$\\
                        & LSTM      & $16.9 \pm 0.2$ & $39.0 \pm 0.4$ & $47.1 \pm 0.2$ & $13.0 \pm 0.2$ & $26.9 \pm 0.3$  & $31.2 \pm 0.2$\\
                        & LSTM attn & $17.0 \pm 0.4$ & $39.4 \pm 0.2$ & $47.5 \pm 0.4$ & $13.6 \pm 0.4$ & $27.1 \pm 0.3$  & $31.4 \pm 0.3$\\
                        & Deepmove  & $19.3 \pm 0.2$ & $43.3 \pm 0.5$ & $52.3 \pm 0.3$ & \underline{$15.5 \pm 0.3$} & $30.1 \pm 0.2$  & $34.9 \pm 0.2$\\
                        & MobTcast  & $\bm{20.2 \pm 0.2}$ & \underline{$45.9 \pm 0.1$} & \underline{$55.5 \pm 0.1$} & $\bm{16.6 \pm 0.2}$ & \underline{$31.7 \pm 0.1$}  & \underline{$36.8 \pm 0.1$}\\
                        & Ours (MHSA)  & $\bm{20.2 \pm 0.3}$ & $\bm{47.0 \pm 0.2}$ & $\bm{57.3 \pm 0.4}$ & $14.9 \pm 0.5$ & $\bm{32.2 \pm 0.3}$ & $\bm{37.6 \pm 0.3}$\\ \midrule

    \multirow{6}{*}{Gowalla}        
                        & 1-MMC     & 10.5 & 19.1 & 20.9 & 9.7 & 14.2 & 15.9 \\
                        & FPMC      & $8.6 \pm 0.4$ & $21.5 \pm 1.1$ & $27.5 \pm 1.3$ & $4.1 \pm 0.2$ & $15.0 \pm 0.7$ & $17.3 \pm 0.8$ \\
                        & LSTM      & $13.2 \pm 0.4$ & $27.7 \pm 0.1$ & $33.6 \pm 0.3$ & $9.4 \pm 0.03$ & $20.1 \pm 0.3$ & $22.8 \pm 0.2$\\
                        & LSTM attn & $13.8 \pm 0.1$ & $29.0 \pm 0.1$ & $35.5 \pm 0.1$ & $10.3 \pm 0.1$ & $21.1 \pm 0.1$ & $23.9 \pm 0.1$\\
                        & Deepmove  & \underline{$14.4 \pm 0.1$} & \underline{$29.7 \pm 0.1$} & \underline{$36.1 \pm 0.1$} & \underline{$11.2 \pm 0.1$} & \underline{$21.7 \pm 0.1$} & \underline{$24.5 \pm 0.1$} \\
                        & MobTcast  & \underline{$14.3 \pm 0.1$} & $28.9 \pm 0.3$ & $34.9 \pm 0.4$ & \underline{$11.1 \pm 0.1$} & $21.3 \pm 0.2$ & $24.0 \pm 0.2$ \\
                        & Ours (MHSA)     & $\bm{15.6 \pm 0.2}$ & $\bm{32.3 \pm 0.2}$ & $\bm{39.2 \pm 0.1}$ & $\bm{12.0 \pm 0.03}$ & $\bm{23.5 \pm 0.2}$ & $\bm{26.6 \pm 0.2}$ \\ \bottomrule

    \end{tabular}

}
\end{table*}

\subsection{Comparing individual and collective models}\label{app:compare}

Designing experiments that ensure a fair comparison between models trained on individual or population data is challenging. We lack a theoretical measure of how many collective patterns are learned by the network and to what degree location visits are caused by personal preference. In the main text, we kept the model architectures identical to compare their number of parameters and performances, which is not ideal as an individual-level model might require less capacity (e.g., fewer layers and heads). However, our experiment suggests that performance variations are insignificant when changing the architecture for individual-level models. These models' total number of parameters is also consistently larger than the one in the collective-level model. Moreover, we note that the training time for the collective-level model is at least one order of magnitude faster than the total training time of the individual-level models. 

In short, we conclude that the model trained on the whole population is more efficient in terms of parameter number and training time, and achieves significantly higher prediction performances for the next location prediction task.

\subsection{Availability of data}\label{app:availability}
Raw data for the GC dataset are not publicly available due to confidentiality agreements with the participants under the European General Data Protection Regulation (GDPR). Raw data for the Geolife dataset is publicly available and can be acquired from the website (\url{https://www.microsoft.com/en-us/research/publication/geolife-gps-trajectory-dataset-user-guide/}).

\end{document}